\documentclass[11pt]{article}          
\makeatletter
\def\@begintheorem#1#2{\trivlist%
 \item[\hskip \labelsep{\sffamily\bfseries #2\ #1}]\itshape}
\newtheorem{teo}{Theorem}[section]
\newtheorem{defi}[teo]{Definition}

\newtheorem{lem}[teo]{Lemma}
\newtheorem{pro}[teo]{Proposition}
\newtheorem{_rem}[teo]{Remark}
\newtheorem{_eje}[teo]{Example}
\newenvironment{rem}{\def\@begintheorem##1##2{\trivlist%
 \item[\hskip\labelsep{\sffamily\bfseries ##2\ ##1}]}\begin{_rem}}{\end{_rem}}

\makeatother

\newenvironment{beweis}{{\em Proof:}}{\hfill $\rule{2mm}{2mm}$
\vspace{3mm}

}

\DeclareMathAlphabet{\Ma}{U}{msa}{m}{n}
\DeclareMathAlphabet{\Mb}{U}{msb}{m}{n}
\DeclareMathAlphabet{\Meuf}{U}{euf}{m}{n}

\def\z#1{\Mb{#1}}
\def\got#1{\Meuf{#1}}

\DeclareSymbolFont{ASMa}{U}{msa}{m}{n}
\DeclareSymbolFont{ASMb}{U}{msb}{m}{n}
\DeclareMathSymbol{\hrist}{\mathord}{ASMa}{"16}
\DeclareMathSymbol{\varkappa}{\mathalpha}{ASMb}{"7B}
\DeclareMathSymbol{\CrPr}{\mathord}{ASMb}{"6F}

\newfont{\EinsFont}{cmr7 scaled 1070}
\def\EINS{{\mathchoice{
 \mbox{\unitlength1cm\begin{picture}(.25,.2)\put(0,0){$1$}%
 \put(0.105,0){{\mbox{\fontfamily{cmr}\upshape\small l}}}\end{picture}}}{%
 \mbox{\unitlength1cm\begin{picture}(.25,.2)\put(0,0){$1$}%
 \put(0.105,0){{\mbox{\fontfamily{cmr}\upshape\small l}}}\end{picture}}}{%
 \mbox{\unitlength1cm\begin{picture}(.18,.15)\put(0,0){$\scriptstyle 1$}%
 \put(0.07,0){{\mbox{\fontfamily{cmr}\upshape\EinsFont l}}}\end{picture}}}{%
 \mbox{\unitlength1cm\begin{picture}(.18,.15)\put(0,0){$\scriptstyle 1$}%
 \put(0.07,0){{\mbox{\fontfamily{cmr}\upshape\EinsFont l}}}\end{picture}}}}}

  \def\al #1.{{\mathcal{#1}}}
  \def\ot #1.{{\got{#1}}}
  \def\C{\Mb{C}}
  \def\N{\Mb{N}}
  \def\R{\Mb{R}}
   
\def\T{\Mb{T}}
\def\DD{\Mb{D}}
\def\TT{\Mb{T}_+}
\def\B{\al O.}
\def\BR{\al B.(\R^4)}

\def\ed{\end{document}}
\def\be{\begin{equation}}
\def\ee{\end{equation}}
\def\bea{\begin{eqnarray}}
\def\eea{\end{eqnarray}}
\def\beaO{\begin{eqnarray*}}
\def\eeaO{\end{eqnarray*}}

\def\g#1#2{\mbox{$\Wort{#1}(#2)$}}
\def\Wort#1{\mbox{\fontfamily{cmr}\selectfont\mdseries\upshape #1}}
\def\In#1{{\fontsize{8pt}{10pt}\selectfont\Wort{#1}}}
\def\KIn#1{{\fontsize{6pt}{8pt}\selectfont\Wort{#1}}}
\newcommand{\bp}{{\breve{p}}} 

\newcommand{\SPgrr}  {\mbox{$\cal P_{+}^{\uparrow}$}}

\newcommand{\UPgr}   {\widetilde{\SPgrr}}
\newcommand{\mudp}   {\mbox{$\;\mu_{0}(\mbox{{\rm d}}p)$}}
\newcommand{\mumdp}  {\mbox{$\;\mu(\mbox{{\rm d}}p)$}}
\newcommand{\Lk}     {\mbox{${\cal C}_{+}^{\circ}$}}
\newcommand{\Lkf}     {\mbox{${\cal C}_{+}$}}
\newcommand{\kSPgr}{\mbox{$\seins_{\kern-.2em\szwei}^{\kern-.1em\sdrei}$}}

\def\ccr#1{\Wort{CCR}(#1)}
\def\car#1{\Wort{CAR}(#1)}
\def\TestO#1{\Wort{C}_0^\infty(#1)}
\def\Test#1{{\cal S}\!\left(#1\right)}

\DeclareMathSymbol{\hsemi}{\mathord}{ASMb}{"6E}
\newcommand{\semi}[2]{\mbox{$#1\kern.1em\hsemi\kern.1em#2$}}
\def\askp#1#2{\mbox{$\LA#1\mbox{\bf ,}\;#2\RA$}}
\def\skp#1#2{\mbox{$\LR#1\mbox{\bf ,}\;#2\RR$}}
\def\LA{\left\langle\bgroup}
\def\LE{\left[\bgroup}
\def\LG{\left\{\bgroup}
\def\LR{\left(\bgroup}
\def\RA{\egroup^{\rule{0mm}{2mm}}\right\rangle}
\def\RE{\egroup^{\rule{0mm}{2mm}}\right]}
\def\RG{\egroup^{\rule{0mm}{2mm}}\right\}}
\def\RR{\egroup^{\rule{0mm}{2mm}}\right)}
\def\Ldummy{\left.\bgroup}
\def\Rdummy{\egroup^{\rule{0mm}{2mm}}\right.}

\def\Kbegin{\begin{equation} \left. \begin{array}{rcl}}
\def\Kend{\end{array} \right\} \end{equation}}
\newenvironment{Klammer}{\Kbegin}{\Kend}
\newcommand{\rep}     {rep\-resenta\-tion}

\title{\bf Conformal covariance of massless free nets}
\author{
 {\sc Fernando Lled\'o}\thanks{On leave from {\em
  Mathematical Institute, University of Potsdam,
  Am Neuen Palais 10, Postfach 601~553,
  D--14415 Potsdam, Germany.} }  \\[2mm] 
 {\footnotesize Max--Planck--Institut f\"ur Gravitationsphysik,}     \\   
 {\footnotesize Albert--Einstein--Institut,}
 {\footnotesize Am M\"uhlenberg 1,}             \\ 
 {\footnotesize D--14476 Golm, Germany.}                        \\
 {\footnotesize lledo@aei-potsdam.mpg.de}}

\date{\today{}}

\begin{document}
\maketitle

\begin{center}
{\sl Dedicated to Hellmut Baumg\"artel on the occasion of his 65th birthday.}
\end{center}
\vspace{.5cm}

\begin{abstract}
In the present paper we review in a fibre bundle context the covariant
and massless canonical representations of the Poincar\'e group as well
as certain unitary representations of the conformal group (in 4 dimensions).
We give a simplified proof of the well--known fact that massless canonical
representations with discrete helicity extend to unitary and irreducible
representations of the conformal group mentioned before. Further we give
a simple new proof that massless free nets for any helicity value are
covariant under the conformal group. Free nets are the result of a 
direct (i.e.~independent of any explicit use of quantum fields) and 
natural way of constructing nets of abstract C*--algebras indexed by open
and bounded regions in Minkowski space that satisfy standard axioms
of local quantum physics. We also give a group theoretical interpretation
of the embedding $\ot I.$ that completely characterizes the free net:
it reduces the (algebraically) reducible covariant representation 
in terms of the unitary canonical ones. Finally, as a consequence of the
conformal covariance we also mention for these models 
some of the expected algebraic properties that are a direct consequence
of the conformal covariance (essential duality, PCT--symmetry etc.).
\end{abstract}

\section{Introduction}\label{Intro}
The birth of massless particles can be traced back to the seminal 
paper \cite{Einstein05} as well as to the most remarkable
part of Einstein's famous principle of special relativity 
also published in 
1905 \cite{Einstein05a} (cf.~also \cite{Einstein09b}):
``{\em Wir wollen diese Vermutung (deren Inhalt im folgenden  
,,Prinzip der Relativit\"at'' genannt werden wird) zur 
Voraussetzung erheben und au{\ss}erdem die mit ihm nur 
scheinbar unvertr\"agliche Voraussetzung einf\"uhren,
da{\ss} sich das Licht im leeren Raume
stets mit einer bestimmten, vom Bewegungszustande des 
emittierenden K\"orpers unabh\"angigen Geschwindigkeit $V$ fortpflanze.}"
Despite their short history
(in comparison with the deeply rooted notion of mass in the physical
literature \cite{bJammer81}) massless particles are related to several 
peculiarities in the analysis of the different branches
in physics where they enter. For example, extrapolating from the principle
above, massless particles inherit a characteristic kinematical behaviour.
This aspect of masslessness is used for instance in the corresponding
collision theory in quantum field theory (henceforth denoted by
QFT): indeed, it is an essential feature of this theory the 
fact that a massless particle (say at the origin) will be for
suitable $t\not=0$ space--like separated from any point in the 
interior of the light--cone (cf.~\cite{Buchholz75,Buchholz77} and
see also \cite{BorchersIn94} for further consequences of the postulate
of maximal speed in classical and quantum physics). 
A different characteristic aspect of masslessness that
will be important in this paper appears in
Wigner's analysis of the unitary irreducible representations of the 
Poincar\'e group \cite{Wigner39} (which is the symmetry group of 
4--dimensional Minkowski spacetime). Indeed, in this analysis
one obtains that the massless
little group $\al E.(2)$ (see (\ref{Delta})) is noncompact, solvable
and has a semi--direct product structure, while the massive little group,
SU(2), satisfies the complementary properties of being compact and simple.
Consequences of these differences will obviously only appear for nonscalar
models, i.e.~in those cases where the corresponding little group
is nontrivially
represented. For example, in order to get discrete helicity values
the solvability and connectedness of $\al E.(2)$ forces to consider only
nonfaithful one--dimensional representations of it, and this fact is 
related to the physical picture characteristic for $m=0$ that the helicity
is relativistic invariant. On the quantum field theoretical side
this aspect appears through the need to
reduce the degrees of freedom of the fibre of the covariant representation
(cf.~Section~\ref{Indu} for the group theoretical definitions and the 
begining of Section~\ref{mfn} for the description of three equivalent 
ways of performing the reduction). A further characteristic feature
of free massless quantum field theoretical models with discrete
helicity is that they are covariant w.r.t.~the conformal group,
i.e.~a bigger symmetry group containing as a
subgroup the original Poincar\'e group with which one starts the 
analysis. In the scalar case (cf.~e.g.~\cite{Kastrup70,Hislop82})
one can argue formally that the space of solutions of the wave--equation
is invariant under the transformation $f\to Rf$, $\big(Rf\big)(x):=
-\frac{1}{x^2}f\big(-\frac{x}{x^2}\big)$, where the relativistic
ray inversion $x\to -\frac{x}{x^2}$
is one of the generating elements of the conformal group.
For higher helicities the conformal covariance of the massless
quantum fields still remains true \cite{Mack69,Hislop88} and due 
to the uniqueness result in 
\cite{Angelopoulos78,Angelopoulos81} ({\em only} the 
unitary irreducible representations of the Poincar\'e group with
$m=0$ and discrete helicity extend within
the same Hilbert space to certain unitary representations
of the conformal group) it is clear that the reduction of the degrees of
freedom mentioned above is an essential feature of the nonscalar models
in order to preserve the conformal group as a symmetry group.
The conformal covariance will have in its turn remarkable structural
consequences for the models. (For a physical interpretation as well
as a historical survey on physical applications of the conformal 
group we refer to \cite{Kastrup62,Todorov78}).

The intention of the present paper is twofold. On the one hand
we review in a fibre bundle context some of the mathematical
peculiarities of the unitary and irreducible representations corresponding
to $m=0$ and discrete helicity (including a simplified proof of the
extension result to a unitary representation of the conformal group). On
the other hand we want to give a simple new proof of the fact that
in QFT massless models with arbitrary helicity are covariant under
the conformal group as well as to apply to these models the important
consequences of this covariance. (Here we treat the helicity values
as a parameter and no special emphasis is laid on the scalar case.)
The simplicity of the proofs mentioned before is partially based on 
the choice of the notion of a {\em free net} in the axiomatic context
of `local quantum physics' (also called algebraic QFT
\cite{bHaag92,Haag64}).
Free nets as considered in \cite{Lledo95,tLledo98} are the result of a
direct and natural
way of constructing nets of abstract C*--algebras indexed by open
and bounded regions in Minkowski space and satisfying Haag--Kastler 
axioms. The construction is based on group
theoretical arguments (concretely on the covariant and canonical
representations of the Poincar\'e group to be introduced in the
following section) and 
standard CAR-- or CCR--theory \cite{ArakiIn87,Manuceau73}. 
In the construction no representation of
the C*--algebra is used and no quantum fields are explicitly needed
and this agrees with the point of view in local quantum
physics that the abstract
algebraic structure should be a primary definition of the theory and
the corresponding Hilbert space representation a secondary 
\cite[Section~4]{pBuchholz99}. In the context of massless models and in
particular in gauge quantum field theory this position is not only an
esthetic one. Indeed, if constraints are present 
in the context of bosonic models the use of 
nonregular representations is sometimes
unavoidable at certain stages of 
the constraint reduction procedure, so that in this frame
one is not always allowed to think of the Weyl elements a `some
sort of exponentiated quantum fields' 
(cf.~\cite{Grundling88c,Grundling88b,pGrundling99}).
Further, the choice of free nets particularly pays off in the massless
case, since here the use of quantum fields unnecessarily complicates the
construction (recall the definition of Weinberg's $2j+1$--fields that
must satisfy the corresponding first--order constraint equation
\cite{Weinberg64a,WeinbergIn65,Hislop88}; the 
necessity of introducing constraints is related to the reduction
of the degrees of freedom of the covariant representation mentioned
above). Finally, we hope that the study
of the mathematical aspects characteristic for massless models
will be useful in the analysis of open problems in mathematical
physics, where masslessness and nontrivial helicity plays a 
significant role (e.g.~in the context of superselection theory,
cf.~\cite{BuchholzIn97}).

The present paper is structured in 5 sections: in the following 
section we review in the general frame of induced representations
on fibre bundles the covariant and canonical representations
of the Poincar\'e group. We will also point out some of the mathematical
differences that appear between the massive and massless canonical
representations. Further, we also consider in this context a method to 
obtain certain unitary representations of the conformal group that
will be needed later. In Section~\ref{mfn} we present the definition
of a massless free net and state some of its properties, for example
they satisfy the Haag--Kastler axioms. The construction is particularly
transparent, because of the use of certain reference spaces, where
the corresponding sesquilinear form is characterized by positive
semidefinite operator--valued functions 
$\beta(\cdot)$ on the mantle of the forward
light--cone $\Lkf$. The corresponding factor Hilbert spaces
(w.r.t.~the degenerate subspace) will carry a representation equivalent
to the unitary irreducible canonical representation with $m=0$
and helicities $\pm\frac{n}{2}$.
In the following section we give a simplified proof of the well--known
fact that the massless Wigner representations mentioned before
extend to certain unitary representations of the conformal group.
For the proof the factor space notation of the previous section
will be useful. Finally, Section~\ref{conseq} shows the covariance
under the conformal group of the massless free nets for any helicity
value. Further, using certain natural Fock states and considering
the corresponding net of von Neumann algebras we are able 
to apply the general results stated in \cite{Brunetti93} for 
conformal quantum field theories to obtain standard algebraic
statements (essential duality, Bisognano--Wichmann Theorem etc.) for
these models.

\section{Induced representations: the Poincar\'e and the conformal group}
\label{Indu}
In the present section we will summarize some results concerning the theory
of induced \rep{s} in the context of fibre bundles. For details and
further generalizations we refer to 
\cite{Asorey85,bSimms68,bSternberg95} and \cite[Section~5.1]{bWarner72}.
We will see below that this general theory beautifully includes all 
representations of the Poincar\'e and the conformal group needed in this
paper. For further aspects of the role played by induced representations
in classical and quantum theory see \cite{Landsman92} and
references cited therein.

Let $\cal G$ be a Lie group that acts transitively on a
C$^\infty$--manifold $M$. Let $u_0\in M$ and 
$\al K._0:=\{g\in \al G.\mid gu_0=u_0\}$ the corresponding 
little group w.r.t.~this action. Then by 
\cite[Theorem~3.2 and Proposition~4.3]{bHelgason78} we have that
$g\al K._0\mapsto gu_0$ characterizes the diffeomorphism
\[
{\al G.}/{\al K._0}\cong \DD :=\{gu_0\mid g\in \al G.\}\,.
\]
In this context we may consider the following principal 
${\cal K}_0$--bundle,  
\begin{equation}
\label{PBundle1}
  {\cal B}_1:= \LR \al G.,\, \Wort{pr}_1,\, \DD \RR.
\end{equation}
$\Wort{pr}_1\colon\ \al G.\to \DD$ denotes the canonical
projection onto the base space $\DD$. Given a representation
$\tau\colon\ {\cal K}_0\to \g{GL}{{\cal H}}$ on the
finite--dimensional Hilbert space ${\cal H}$, one can construct the 
associated vector bundle 
\begin{equation}
\label{PBundle2}
 {\cal B}_2(\tau):= 
    \LR \al G.\times_{\KIn{${\cal K}_0$}}{\cal H},\;\Wort{pr}_2,\;\DD \RR.
\end{equation}
The action of $\cal G$ on $M$
specifies the following further actions on $\DD$ and on 
$\al G.\times_{\KIn{${\cal K}_0$}}{\cal H}$: for 
$g,g_0\in {\cal G}$, $v\in \al H.$, put
\begin{equation}\label{PMActions}
\left. \begin{array}{rcl}
{\cal G}\times \DD &\longrightarrow& \DD , \kern2.5cm 
                            g_0\; \Wort{pr}_1(g):=\Wort{pr}_1(g_0g) \\
{\cal G}\times \LR \al G.\times_{\KIn{${\cal K}_0$}}{\cal H} \RR 
           &\longrightarrow& \al G.\times_{\KIn{${\cal K}_0$}}{\cal H}, \kern1.82cm 
           g_0\; [g,v]:=[g_0g,v]\,, \end{array} \right\}
\end{equation}
where $[g,v]=[gk^{-1},\tau(k)v]$, $k\in\al K._0$, denotes 
the equivalence class characterizing a point in the 
total space of the associated bundle.
Finally we define the (from $\tau$) induced representation of ${\cal G}$ on
the space of sections
of the vector bundle ${\cal B}_2$, which we denote by
$\Gamma (\al G.\times_{\KIn{${\cal K}_0$}}{\cal H})$:
let $\psi$ be such a section and for
$g\in {\cal G}$, $p\in \DD$:
\begin{equation}
\label{FbInd}
 \LR T(g)\psi\RR\!(p):=g\; \psi\!\LR g^{-1} p\RR .
\end{equation}

\begin{rem}\label{HFunct}
We will now present two ways of rewriting the preceding induced
representation in (for physicists more usual)
terms of vector--valued functions.
\begin{itemize}
\item[(i)] The more standard one consists of choosing a section
 $s\colon\ \DD\to \al G.$ of the principal ${\cal K}_0$--bundle 
 ${\cal B}_1$. Now for $\psi\in
 \Gamma (\al G.\times_{\KIn{${\cal K}_0$}}{\cal H})$ we put
 $\psi(p)=[s(p),\varphi(p)]$, $p\in\DD$, for a suitable function
 $\varphi\colon\ \DD\to\al H.$ and we may rewrite the induced \rep{} as 
 \begin{equation}\label{VfInd}
  \LR T(g)\varphi\RR\!(p)=\tau\!
  \LR s(p)^{-1}g\, s\!\LR g^{-1}p\RR \RR \varphi\!\LR g^{-1} p\RR,
 \end{equation}
 where it can be easily seen that 
 $s(p)^{-1}g\, s\!\LR g^{-1}p\RR\in {\cal K}_0$.

\item[(ii)] A second less well known way of transcribing the induced
 \rep{} (\ref{FbInd}) is done by means of a 
 mapping $J\colon\ \al G.\times \DD\to \g{GL}{\al H.}$ that satisfies
 \begin{eqnarray}
 \label{J1} J(g_1g_2,p) 
      &=& J(g_1,g_2 p)\,J(g_2,p)\,,\quad g_1,g_2\in\al G.\,,\;p\in\DD\\
 \label{J2} J(e,p)&=&\EINS\,,\quad\mbox{where $e$ is the unit in $\al G.$}\\ 
 \label{J3} J(k,u_0) &=& \tau(k) \,,\quad k\in\al K._0\,.
 \end{eqnarray}
 Note that by (\ref{J1}) the l.h.s.~of Eq.~(\ref{J3}) is indeed a \rep{}
 of $\al K._0$. Now for $\psi\in
 \Gamma (\al G.\times_{\KIn{${\cal K}_0$}}{\cal H})$ and a suitable 
 function $\varphi\colon\ \DD\to\al H.$ we may put
 $\psi(p)=[g,J(g,u_0)^{-1}\,\varphi(p)]$, $g\in\al G.$ and 
 $\Wort{pr}_1(g)=p\in\DD$, which is a consistent expression w.r.t.~the
 equivalence classes in $\al G.\times_{\KIn{${\cal K}_0$}}{\cal H}$:
 indeed, using (\ref{J1}) and (\ref{J3}) above we have for any
 $k\in\al K._0$
\[
 \psi(p)=[g,J(g,u_0)^{-1}\,\varphi(p)]
        =[gk^{-1},\tau(k)\,J(g,u_0)^{-1}\,\varphi(p)]
        =[gk^{-1},J(gk^{-1},u_0)^{-1}\,\varphi(p)]\,.
\]
 From this we may rewrite the induced \rep{} as
\begin{equation}\label{KerJ}
 \LR T(g_0) \varphi\RR\!(p)=J\!\LR g_0^{-1},p\RR^{-1} 
     \varphi(g_0^{-1} p) \,,\quad g_0\in\al G.\,,\; p\in\DD\,.
\end{equation}
 Using for example (\ref{J1})--(\ref{J3}) above it can be directly checked
 that $T$ is indeed a \rep{}. The present analysis in terms of the 
 mapping $J$ will be useful later in the context of the conformal
 group (cf.~\cite[Section~I.4]{Jakobsen77}). 
\end{itemize}
\end{rem}

Note that till now we have not specified any structure on the sections 
$\Gamma (\al G.\times_{\KIn{${\cal K}_0$}}{\cal H})$ (or on the
set of $\al H.$--valued functions). In the following we will apply
the preceding general scheme to the Poincar\'e and the conformal group
and will completely fix the structure of the corresponding
representation spaces. We will also give regularity conditions on the 
section $s$ considered in part (i) above.

\subsection{The Poincar\'e group:}\label{thep}
We will specify next the so--called covariant and canonical
representations of the Poincar\'e group. They will play a fundamental 
role in the definition of the free net in the next section. Besides the 
references mentioned at the begining of this section we refer also to
\cite{Barut72,bBarut80,bMackey76,Wigner39} as well as
\cite[Section~2.1]{Landsman95b}.

\paragraph{Covariant representations:}
In the general analysis considered above let 
$\al G.:=\semi{\g{SL}{2,\C}}{\R^4}=\UPgr$ be the universal covering
of the proper orthocronous component of the Poincar\'e group. 
It acts on $M:=\R^4$ in the usual way 
$(A,a)\,x:=\Lambda_{A}x+a$, $(A,a)\in\semi{\g{SL}{2,\C}}{\R^4}$, 
$x\in\R^4$, where $\Lambda_{A}$ is the Lorentz transformation 
associated to $\pm A\in\g{SL}{2,{\C}}$ which
describes the action of $\g{SL}{2,\C}$ on $\R^4$
in the last semidirect product. Putting now $u_0:=0$ gives
$\al K._0=\semi{\g{SL}{2,\C}}{\{0\}}$, 
${{\cal G}}/{\LR\semi{\g{SL}{2,\C}}{\{0\}}\RR}\cong
\R^4$, and the principal $\g{SL}{2,\C}$--bundle is in this case
${\cal B}_1:=({\cal G},\; \Wort{pr}_1,\; \R^4)$.
As inducing representation we use the finite--dimensional 
irreducible \rep{s} of $\g{SL}{2,{\C}}$ acting on the spinor 
space ${\cal H}^{\KIn{$(\frac{j}{2},\frac{k}{2})$}}
:=\Wort{Sym}\Big(\mathop{\otimes}\limits^j \z{C}^2\Big)
\otimes \Wort{Sym} \Big( \mathop{\otimes}\limits^k \z{C}^2 \Big)$ 
(cf.~\cite{bStreater89}):
i.e.~$\tau^{\KIn{(cov)}}(A,0):=D^{\KIn{$(\frac{j}{2},\frac{k}{2})$}}(A)
=\Big(\mathop{\otimes}\limits^j A\Big)
  \otimes \Big( \mathop{\otimes}\limits^k \overline{A} \Big)$, 
$(A,0)\in \semi{\g{SL}{2,\C}}{\{0\}}$.
From this we have (if no confusion arises we will omit in the 
following the index \In{$(\frac{j}{2},\frac{k}{2})$} in $D(\cdot)$ and 
in $\cal H$),
\begin{equation}
\label{CovB2}
  {\cal B}_2\!\LR\tau^{\KIn{(cov)}}\RR:= \LR 
                   {\cal G}\times_{\KIn{SL$(2,{\C})$}}{\cal H},\;
                   \Wort{pr}_2,\; \R^4\RR.
\end{equation}
Recalling Remark~\ref{HFunct}~(i) we specify a global continuous 
section $s$ of ${\cal B}_1$ (i.e.~${\cal B}_1$ is a trivial bundle):
\[
 s\colon\ \R^4\longrightarrow {\cal G},\qquad s(x):=(\EINS,x)\in
 \semi{\g{SL}{2,{\C}}}{\R^{4}}={\cal G}\,.
\]
Note that since $\tau^{\KIn{(cov)}}$ is {\em not} a unitary representation
and since we want to relate the following so--called
{\em covariant \rep{}} with the irreducible and
unitary canonical ones presented below, it is enough to define 
$T$ on the space of $\cal H$--valued Schwartz functions
$\Test{\R^{4},{\cal H}}$
\begin{equation}
\label{VfCov}
 \LR T(g)f\RR\!(x):= D(A)\, f\!\LR\Lambda_{A}^{-1}
 (x-a)\RR,\quad f\in \Test{\R^{4},{\cal H}}\,,
\end{equation}
where we have used that $s(x)^{-1}\,(A,a)\,s\!\LR(A,a)^{-1}x\RR =(A,0)$, 
$(A,a)\in {\cal G}$.
$T$ is an algebraically reducible \rep{} even if the
inducing \rep{} $\tau^{\KIn{(cov)}}$ is irreducible.

\begin{rem}
In \cite{tLledo98,pLledo99b} it is shown that the covariant \rep{}
is related with the covariant transformation character of quantum 
fields. Thus a further reason for considering this \rep{}
space is the fact that in the heuristic picture we want to smear
free quantum fields with test functions in $\Test{\R^{4},{\cal H}}$.
\end{rem}

\paragraph{Canonical representations:}
Next we will consider unitary and irreducible canonical representations 
of $\UPgr$ and in particular specify the massless ones with discrete
helicity. We will apply in this case Mackey's theory of 
induced \rep{s} of regular semidirect products, 
where each subgroup is locally compact and one of them 
abelian \cite{bMackey76,bSimms68,bBarut80}.

First note that in the general context of the begining of this section 
if $\tau$ is a unitary \rep{} of $\al K._0$ on $\al H.$, then
$\Gamma (\al G.\times_{\KIn{${\cal K}_0$}}{\cal H})$
turns naturally into a Hilbert space. Indeed, the fibres
$\Wort{pr}_2^{-1}(p)$, $p\in \DD$, inherit
a unique (modulo unitary equivalence) Hilbert space structure 
from ${\cal H}$. Assume further that $\DD$ allows 
a ${\cal G}$--invariant  measure $\mu$. (The following construction goes also
through with little modifications if we only require the existence on $\DD$
of a quasi--invariant measure w.r.t.~${\cal G}$.) Then
$\Gamma ( \al G.\times_{\KIn{${\cal K}_0$}}{\cal H})$ is the Hilbert
space of all measurable sections $\psi$ of ${\cal B}_2(\tau)$ 
that satisfy,
\[
 \askp{\psi}{\psi}=\int\limits_\DD\askp{\psi(p)}{\psi(p)}_p \mumdp <\infty,
\]
where $\langle\cdot,\cdot\rangle_p$ denotes the scalar product on the Hilbert
space $\Wort{pr}_2^{-1}(p)$, $p\in \DD$, and the induced \rep{}
given in Eq.~(\ref{FbInd}) is unitary on it.

Put now $\al G.:=\g{SL}{2,\C}$ which acts on $\widehat{\R}^4$
by means of the dual action canonically given by the semidirect product 
structure of $\UPgr$.
It is defined by $\widetilde{\gamma}\colon\ \g{SL}{2,\C} 
\rightarrow \Wort{Aut}\,\widehat{\R}^4$, $\chi\in
\widehat{\R}^4$, and $(\widetilde{\gamma}_{\KIn{$A$}}\chi)(a)
:=\chi(\Lambda_{A}^{-1} (a))$, $A\in \g{SL}{2,\C}$, $a\in \R^4$.
For $\chi\in\widehat{\R}^4$ fixed the corresponding little and 
isotropy subgroups are defined respectively by 
\[
 {\cal G}_\chi:=\LG A\in \g{SL}{2,\C} \mid 
 \widetilde{\gamma}_{\KIn{$A$}}\chi=\chi \RG \,,\;\;
 {\cal I}_{\chi}:=\semi{{\cal G}_\chi}{\R^4} 
 \quad\Wort{and note that}\quad
 {\UPgr}/{{\cal I}_{\chi}}\cong {{\cal G}}/{{\cal G}_{\chi}}
 \cong\DD\,.
\]
We have now the principal ${\cal I}_{\chi}$--bundle and the associated
bundle given respectively by
\[
  {\cal B}_1:=\LR\UPgr,\;\Wort{pr}_1,\;\DD\RR\quad\Wort{and}\quad
  {\cal B}_2\!\LR\tau^{\KIn{(can)}}\RR:= 
          \LR\UPgr\times_{\KIn{${\cal I}_{\chi}$}}{\cal H},
          \;\Wort{pr}_2,\;\DD\RR ,
\]
where $\tau^{\KIn{(can)}}$ is a unitary \rep{} of
${\cal I}_{\chi}$ on ${\cal H}$.
If $\tau^{\KIn{(can)}}$ is irreducible, then the corresponding induced
\rep{}, which is called the {\em canonical representation}, is irreducible.
Even more, every irreducible \rep{} of ${\cal G}$ is obtained (modulo unitary
equivalence) in this way. Recall also that the canonical \rep{} is 
unitary iff $\tau^{\KIn{(can)}}$ is unitary.

To specify massless representations with discrete helicity we choose
a character $\chi_{\bp}$, $\bp:=(1,0,0,1)\in\Lkf$ 
(the mantle of the forward light cone), i.e.
$\chi_{\bp}(a)=e^{-i\bp a}$, $a\in\R^4$ and $\bp\, a$ means the
Minkowski scalar product. A straightforward computation shows that
the isotropy subgroup is given by
${\cal I}_{{\chi}_{\bp}}=\semi{{\cal E}(2)}{\R^4}$, where
\begin{equation}
\label{Delta}
{\cal E}(2) := \left\{
            \left(\kern-1.5mm \begin{array}{cc}
            e^{\frac{i}{2}\theta} & e^{-\frac{i}{2}\theta}\ z               \\
                                0 & e^{-\frac{i}{2}\theta}
            \end{array} \kern-1.5mm\right) \in \g{SL}{2,\C}
            \mid \theta\in[0,4\pi),\; z\in\C
            \right\}.
\end{equation}
The little group ${\cal E}(2)$ is noncompact and since its commutator
subgroup is already abelian it follows that ${\cal E}(2)$ is 
solvable. Further, it has again the structure of a semidirect product. 
(In contrast with this fact
we have that the massive little group $\g{SU}{2}$ is compact and
simple.) Since ${\cal E}(2)$ is a connected and solvable Lie group we know
from Lie's Theorem (cf.~\cite{bBarut80}) that the only
finite--dimensional irreducible \rep{s} are 1--dimensional, 
i.e.~${\cal H}:=\C$. Therefore in order to induce irreducible 
and unitary \rep{s} of the whole group that describe discrete helicity
values we define
\begin{equation}
\label{Sigma0Can}
 \tau^{\KIn{(can)}}(L,a):=e^{-i\bp a}\, \LR e^{\frac{i}{2}\theta}\RR^n,
\end{equation}
where $(L,a)\in \semi{{\cal E}(2)}{\R^4}={\cal I}_{{\chi}_{\bp}}$, 
$n\in \N$. Note that this \rep{} is not faithful. Indeed, 
the normal subgroup $\left\{
      \left(\kern-1.5mm \begin{array}{cc}
          1 & z               \\
          0 & 1
  \end{array} \kern-1.5mm\right)\mid  z\in\C\right\}$
is trivially represented (see also \cite[Section~II]{Weinberg64a}). 
Some authors associate this subgroup to certain gauge degrees of 
freedom of the system (e.g.~\cite{Han81,Kim87,Shnerb94}).
We consider next the bundles,
\[
{\cal B}_1^{\KIn{(can)}} := \LR \UPgr,\; \Wort{pr}_1,\; 
                              \Lkf \RR \quad\Wort{and}\quad
{\cal B}_2\!\LR\tau^{\KIn{(can)}}\RR
                     := \LR \UPgr\times_{\KIn{${\cal I}_{{\chi}_{\bp}}$}}{
                              \!\C},\; \Wort{pr}_2,\; \Lkf \RR,
\]
where we have used the diffeomorphism  
${\UPgr}/{{\cal I}_{{\chi}_{\bp}}}\cong\Lkf$
between the factor space and the mantle of the forward 
light--cone. We denote by $\mudp$ the corresponding invariant measure on 
$\Lkf$.

In contrast with the massive case the bundle ${\cal B}_1^{\KIn{(can)}}$
has no global continuous section. This fact is based on the comparison of 
different homotopy groups that can be associated with the bundle
${\cal B}_1^{\KIn{(can)}}$ \cite{Boya74}. Nevertheless, 
we can specify a measurable section considering a continuous one
in a chart that does not include the set $\{ p\in\Lkf\mid  p_3=-p_0\}$ 
(which is of measure zero w.r.t.~$\mudp$).
Putting $\Lk:=\Lkf\setminus \{ p\in\Lkf\mid  p_3=-p_0\}$
a (local) continuous section
is given explicitly by
\begin{equation}
\label{Section0}
 s\colon\ \Lk\longrightarrow \UPgr,\qquad s(p):=( H_p,\,0 )\in
             \semi{\g{SL}{2,{\C}}}{\R^{4}}=\UPgr,
\end{equation}
where
\begin{equation}
\label{SolHp}
    H_{p} := \frac{1}{\sqrt{2p_{0}(p_{0}+p_{3})}}
       \left(\kern-1.5mm \begin{array}{cc}
       -\sqrt{p_{0}}\ (p_{0}+p_{3})  &
       \displaystyle \phantom{-}\frac{p_{1}-ip_{2}}{\sqrt{p_{0}}}           \\
       -\sqrt{p_{0}}\ (p_{1}+ip_{2}) &
       \displaystyle -\frac{p_{0}+p_{3}}{\sqrt{p_{0}}}^{}
       \end{array} \kern-1.5mm\right).
\end{equation}
Recall that the $H_p$--matrices satisfy the equation
\begin{equation}
\label{DefHp}
H_{p}^{} \left(\kern-1.5mm \begin{array}{cc} 2\kern-1.6mm &0 \\ 
           0\kern-1.6mm &0 \end{array} \kern-1.5mm\right)
H_{p}^{*} = P, \quad \mbox{where}\quad
P = \left(\kern-1.5mm \begin{array}{cc}
    p_{0}+p_{3}  & p_{1}-ip_{2} \\
    p_{1}+ip_{2} & p_{0}-p_{3}
    \end{array} \kern-1.5mm\right)
  = p_0\sigma_0+\sum \limits_{i=1}^3 p_{i} \sigma_{i},
\end{equation}
where $\sigma_{\mu}$, $\mu=0,1,2,3$, are the unit and the Pauli
matrices and we have used the vector space isomorphism between $\R^4$
and $\g{H}{2,\C}:=\{ P\in \Wort{Mat}_2(\C)\mid P^*=P\}$ given by
$ \R^4\ni p:=(p_0,p_1,p_2,p_3)\mapsto P$.

If we consider the section in Eq.~(\ref{Section0}) fixed, then we have on 
$\mathrm{L}^2(\Lkf,\C,\mudp)$ the canonical massless representations 
(cf.~Eq.~(\ref{VfInd}))
\begin{equation}
\label{Vf0Can}
 \LR U_{\pm}(g)\varphi\RR\!(p)=e^{-ipa}\LR 
                        e^{\pm\frac{i}{2}\theta(A,p)}\RR^n\varphi (q),
\end{equation}
where $g=(A,a)\in \semi{\g{SL}{2,\C}}{\R^4}$,  
$n\in \N$, $q:=\Lambda_{A}^{-1}p$ and for 
$A=\left(\kern-1.5mm \begin{array}{cc}
          a \kern-2mm& b               \\
          c \kern-2mm& d
  \end{array} \kern-1.5mm\right)\in \g{SL}{2,\C}$ we compute
\[
  e^{-\frac{i}{2}\theta(A,p)}:=\LR H_p^{-1} A\, H_q\RR_{22}=
  \frac{-b(p_{1}+ip_{2})+d(p_{0}+p_{3})}
              {|-b(p_{1}+ip_{2})+d(p_{0}+p_{3})|}\,.
\]
$U_{\pm}$ are unitary w.r.t.~usual L$^2$--scalar product, satisfy the 
spectrum condition and the helicity of the model carrying 
one of these representations is $\pm\frac{n}{2}$.

\subsection{The conformal group:}\label{subconf}
We will consider first some standard facts concerning the conformal 
group \cite[Appendix]{tHislop84},\cite{PetkovaIn89,Todorov78,Jakobsen77}.
We will describe later a technique to define a unitary
\rep{} of $\g{SU}{2,2}$, by means of the mapping 
$J$ considered in Remark~\ref{HFunct}~(ii). 
These results are a variation of the notion of reproducing 
kernel for which we refer to \cite{KunzeIn67,Jakobsen77,Carey77c}
and will be useful in order to extend the massless 
canonical \rep{s} to unitary \rep{s} of the conformal group. 

The group
\be\label{SU22}
 \g{SU}{2,2}:= \{ g\in\Wort{Mat}_4(\C)\mid \Wort{det}\,g=1\; \Wort{and}\;
                g\,\zeta\, g^*=\zeta \}\;,\quad\Wort{where}\;
        \zeta:= \left(\kern-1.5mm \begin{array}{cc}
                      0 \kern-1mm &  -i\EINS               \\
                  i\EINS\kern-1mm &  0
                 \end{array} \kern-1.5mm\right)\,,
\ee
is the fourfold covering of the conformal group in Minkowski space.
Using $A,B,C,D \in\Wort{Mat}_2(\C)$ we have that 
$g=\left(\kern-1.5mm \begin{array}{cc}
                      A \kern-2mm &  B  \\
                      C \kern-2mm &  D
                 \end{array} \kern-1.5mm\right)\in\g{SU}{2,2}$ 
iff $\mathrm{det}\,g=1$ and 
\begin{Klammer}\label{ABCD}
   AB^* = BA^*  \quad    & & \quad C^*A=  A^*C\\
   CD^* = DC^*  \quad &\Wort{or equivalently}& \quad B^*D = D^*B \\
   AD^*-BC^*=\EINS \quad & & \quad A^*D-C^*B=\EINS    \,.
\end{Klammer}

Further, we write the natural action of $\g{SU}{2,2}$ on the forward tube
\[
 \T_+:=\g{H}{2,\C}+i\g{H$_+$}{2,\C}\cong
          \R^4+i\,{\cal V}_+\,,\quad
           \g{H$_+$}{2,\C}:=\{ P\in \g{H}{2,\C} \mid
        \Wort{det}\,P>0\,,\,
        \Wort{Tr}\,P>0\}
\]
as follows:
\be\label{CAction}
 g Z:=(AZ+B)\,(CZ+D)^{-1}\,,\quad Z=X+iY\in\T_+\,.
\ee
Finally, the Poincar\'e group, the dilations and the special conformal
transformations can be recovered as subgroups of $\g{SU}{2,2}$. In 
particular we will need later 
\be\label{SubPoin}
      \UPgr = \LG\left(\kern-1.5mm \begin{array}{cc}
                      A \kern-2mm &  B (A^*)^{-1} \\
                      0 \kern-2mm &  (A^*)^{-1}
                 \end{array} \kern-1.5mm\right)
              \mid A\in\g{SL}{2,\C}\,,\;B\in\g{H}{2,\C}\RG
              \subset \g{SU}{2,2}\,.
\ee

With this action in mind recall the general situation 
concerning induced representations at the begininng of this section 
and put now $\al G.:=\g{SU}{2,2}$,
$M:=\TT$ and $u_0:=i\EINS\in\T_+$, so that from the action
given in Eq.~(\ref{CAction}) we get \cite[Section~3]{Jakobsen77}
\[
  {\cal K}_0:=\{g\in \g{SU}{2,2}\mid gi\EINS =i\EINS\}
       = \LG \left(\kern-1.5mm \begin{array}{cc}
             A \kern-2mm &  B \\
            -B \kern-2mm &  A 
        \end{array} \kern-1.5mm\right)\in \g{SU}{2,2}\RG
\quad\Wort{and}\quad {\g{SU}{2,2}}/{{\cal K}_0}\cong\TT\,.
\]

Suppose now that there exists a Hilbert space $\ot H.$ with scalar
product $\langle\cdot,\cdot\rangle$ (we will identify later $\ot H.$
with the representation Hilbert space of the massless canonical
representations) 
and that we may use $\TT$ and $\al H.$ to parametrize a total set
\[
 {\got H}_{tot}:=\{ K_{z,v}\mid z\in\TT, v\in\al H.\}\subset {\got H}\,.
\]
\begin{lem}\label{TConfUnit}
If the scalar product satisfies on ${\got H}_{tot}$ the property 
\be\label{SkpCov}
 \askp{K_{gz_1,v_1}}{K_{gz_2,v_2}}
  =\askp{K_{z_1,\,J(g,z_1)^*v_1}}{K_{z_2,\,J(g,z_2)^*v_2}}\,,\quad
   z_1,z_2\in\TT\,,\;v_1,v_2\in\al H.,
\ee
for all $g\in\g{SU}{2,2}$, then the \rep{} defined on ${\got H}_{tot}$ by
\[
 V(g)K_{z,v}:=K_{gz,\,(J(g,z)^{-1})^*v}\in {\got H}_{tot}
\]
extends to a unitary \rep{} within $\got H$.
\end{lem}
\begin{beweis}
First of all note that on ${\got H}_{tot}$ the relation
$V(g_1g_2)=V(g_1)V(g_2)$, $g_1,g_2\in\g{SU}{2,2}$, holds. Indeed, 
using Eq.~(\ref{J1}) we have 
\[
 V(g_1g_2)K_{z,v}=K_{g_1g_2z,\,(J(g_1g_2,z)^{-1})^*v}
     =K_{g_1g_2z,\,(J(g_1,g_2z)^{-1})^*\,(J(g_2,z)^{-1})^*v}
     =V(g_1)\,(V(g_2)K_{z,v}).
\]
We can also easily check the isometry property on 
$\Wort{span}\,{\got H}_{tot}$, which by
assumption is dense in $\got H$. For 
$\sum\limits_{l=1}^L\lambda_l\,K_{z_l,v_l}$,
$\sum\limits_{m=1}^M\lambda_m'\,K_{z_m',v_m'}\in
\Wort{span}\,{\got H}_{tot}$ and extending by linearity
the above definition we have
\begin{eqnarray*}
 \lefteqn{\!\!\!\!\!\!\!\!\!\!\!\!\!\!\!\!\!\!\!\!\!\!\!\!\!\!
       \askp{V(g)\LR\sum\limits_{l=1}^L\lambda_l\,K_{z_l,v_l}\RR}{
       V(g)\LR\sum\limits_{m=1}^M\lambda_m'\,K_{z_m',v_m'}\RR}} \\
   &=& \sum_{l,m}\; \overline{\lambda_l}\,\lambda_m'
    \askp{K_{gz_l,(J(g,z_l)^{-1})^*v_l}}{K_{gz_m',(J(g,z_m')^{-1})^*v_m'}}\\ 
   &=& \sum_{l,m}\; \overline{\lambda_l}\,\lambda_m'
    \askp{K_{z_l,J(g,z_l)^*(J(g,z_l)^{-1})^*v_l}}{
          K_{z_m',J(g,z_m')^*(J(g,z_m')^{-1})^*v_m'}}\\ 
   &=& \askp{\sum\limits_{l=1}^L\lambda_l\,K_{z_l,v_l}}{
             \sum\limits_{m=1}^M\lambda_m'\,K_{z_m',v_m'}}\,.
\end{eqnarray*}
We can therefore extend isometrically $V(g)$ to a unitary \rep{}
on the whole $\got H$.
\end{beweis}

\section{Massless free nets}\label{mfn}

In the following we will briefly review with some modifications 
and improvements the massless free net construction
presented in \cite[Part~B]{Lledo95}. 
The fundamental object that characterizes a free net is the linear
embedding $\ot I.$ that intertwines between the covariant and the 
canonical representation. The free net will be called
massive or massless depending if the canonical representation 
corresponds to $m>0$ resp.~$m=0$.
Now a typical feature of massless models with
helicity $\not= 0$ is the fact that the embeddings must reduce the degrees
of freedom on the fibres of the corresponding associated bundles. Indeed,
as a consequence of the fact that $\al E.(2)$ is solvable 
we have that the fibres
of ${\cal B}_2(\tau^{\KIn{(can)}})$ are 1--dimensional, while
the fibres of ${\cal B}_2(\tau^{\KIn{(cov)}})$ are at least
2--dimensional if one chooses a nontrivial inducing \rep{} 
$\tau^{\KIn{(cov)}}$. With other words, if the models describe
nontrivial helicity, then some further restriction must be performed 
on the fibres in order to 
reduce the covariant \rep{} to the unitary and irreducible canonical one. 
There are at least three ways to perform the mentioned reduction that
will produce isomorphic nets of C*--algebras:
\begin{itemize}
\item[(i)] One possibility that will be considered next is to rewrite the 
 massless canonical representation in a for us much more convenient way.
 Using certain natural reference spaces with a semidefinite sesquilinear form 
 characterized by an positive semidefinite
 operator--valued function $\beta(\cdot)$, the 
 reduction is done passing to the factor spaces that can be
 canonically constructed from the degeneracy subspaces of the sesquilinear 
 form. 
\item[(ii)] A second possibility is to consider other type of embeddings 
 that map the Schwartz
 test functions to the space of solutions of the corresponding
 massless relativistic wave equations. Here the reduction is done by means 
 of certain invariant (but not reducing) projections on the spinor space
 $\al H.$ (cf.~\cite{tLledo98,pLledo99b}).

\item[(iii)] Finally, one can also perform the mentioned
 reduction for the bosonic models at the C*--level
 by the constraint reduction procedure of Grundling
 and Hurst \cite{Grundling85}. In this context the constraints can be
 defined as the Weyl elements associated to the degenerate subspace of 
 part (i) (cf.~\cite{Lledo97}) and the constraint reduction here is similar
 to the second stage of reduction of the Gupta--Bleuler model 
 considered in \cite[Theorem 5.14]{pGrundling99}.
\end{itemize}

The essence of the following construction is the fact that for each 
$p\in\Lkf$ the nonnegative, selfadjoint matrix
$P^{\KIn{$\dagger$}} = \frac12 \left(\kern-1.5mm \begin{array}{cc}
                      p_{0}-p_{3}  & -p_{1}+ip_{2} \\
                      -p_{1}-ip_{2} & p_{0}+p_{3}
                      \end{array} \kern-1.5mm\right)$
has the eigenvalues $p_0$ and $0$:
\begin{equation}
\label{EigenWert0}
P^{\KIn{$\dagger$}}\, \left(\kern-1.5mm \begin{array}{c}
                      -p_1 + ip_2 \\
                       p_0 + p_3 
                      \end{array} \kern-1.5mm\right)
         = p_0\,\left(\kern-1.5mm \begin{array}{c}
                      -p_1 + ip_2 \\
                       p_0 + p_3 
                      \end{array} \kern-1.5mm\right)   
\quad\Wort{and}\quad
P^{\KIn{$\dagger$}}\, \left(\kern-1.5mm \begin{array}{c}
                       p_1 - ip_2 \\
                       p_0 - p_3 
                      \end{array} \kern-1.5mm\right)
         = 0 \,\left(\kern-1.5mm \begin{array}{c}
                        p_1 - ip_2 \\
                        p_0 - p_3 
                      \end{array} \kern-1.5mm\right)\,. 
\end{equation}
That $P^{\KIn{$\dagger$}}$ has the eigenvalue 0 is a
typical feature of massless \rep{s}, since for the massive ones the
corresponding matrix $P^{\KIn{$\dagger$}}$ is strictly positive for any
$p$ on the positive mass shell (see \cite[Part A]{Lledo95}).
Now each function $\varphi\colon\ 
\Lkf\setminus \{p\in\Lkf\mid |p_3|=p_0 \}\longrightarrow \z{C}^2$
can be decomposed pointwise into a sum of the eigenvectors above
(recall that $\{p\in\Lkf\mid |p_3|=p_0 \}$ is of measure 0 
w.r.t~$\mu_0$),
\begin{equation}\label{Decomposition}
 \varphi(p)=\left(\kern-1.5mm \begin{array}{c}
                      -p_1 + ip_2 \\
                       p_0 + p_3 
                      \end{array} \kern-1.5mm\right)\;\alpha_+(p)
           + \,\left(\kern-1.5mm \begin{array}{c}
                       p_1 - ip_2 \\
                       p_0 - p_3 
                      \end{array} \kern-1.5mm\right)\;\alpha_0(p)\,,
\end{equation}
for suitable $\z{C}$--valued functions $\alpha_+$, $\alpha_0$. 
Further, the matrix $P^{\KIn{$\dagger$}}$ is a natural 
object from the point of view
of representation theory of the Poincar\'e group. It is straightforward to
show that $P^{\KIn{$\dagger$}}=\left(H_p^{-1}\right)^* 
         \left(\kern-1.5mm\begin{array}{cc} 
              0 \kern-1.6mm & 0             \\
              0 \kern-1.6mm & 1
         \end{array}\kern-1.5mm\right) H_p^{-1}
         =\frac12 \left( p_{0}\sigma_{0}-
          \sum\limits_{i=1}^{3}p_{i}\sigma_{i}\right)$, 
where the matrices $H_p\in\g{SL}{2,\z{C}}$, $p\in\Lk$, are given   
in Eq.~(\ref{SolHp}). 
 
The sesquilinear forms to be defined next are characterized
by the following positive semidefinite operator--valued functions:
put for $p\in\Lkf$ and $n\in\z{N}$,
\[ 
\beta_+(p) := D^{\KIn{$(0,\frac{n}{2})$}}(P^{\KIn{$\dagger$}})=
              \mathop{\otimes}\limits^n \overline{P^{\KIn{$\dagger$}}} 
 \quad\Wort{and}\quad
\beta_-(p) := D^{\KIn{$(\frac{n}{2},0)$}}(P^{\KIn{$\dagger$}}) =
              \mathop{\otimes}\limits^n P^{\KIn{$\dagger$}} \,.
\]
$\beta_{\pm}(p)$ act on ${\cal H}^{\KIn{$(0,\frac{n}{2})$}}$ 
resp.~${\cal H}^{\KIn{$(\frac{n}{2},0)$}}$.
Define then for $\varphi,\psi$ a pair of ${\cal H}$--valued measurable
functions the sesquilinear forms
\begin{equation}
 \label{sesquilinear5}
 \askp{\varphi}{\psi}_{\beta_{\pm}}:= \int\limits_{\Lkf}
 \skp{\varphi(p)}{\beta_{\pm}(p)\ \psi(p)}_{\cal H} \mudp, 
\end{equation}
and from this consider the sets
\begin{equation}
 {\got H}_{n,\pm}:= \LG\varphi\colon\ \Lk\longrightarrow {\cal H}\mid 
 \varphi\; \Wort{is measurable and}\quad 
 \langle\varphi ,\varphi\rangle_{\beta_{\pm}}< \infty \RG \,.
\end{equation}
For $\varphi_{\pm}\in{\got H}_{n,\pm}$ we define also the \rep{s}:
\begin{eqnarray}
  \LR V_1(g)\kern.2em\varphi_+ \RR(p)
    &:=& e^{-ipa}\ D^{\KIn{$(0,\frac{n}{2})$}}(A)\ \varphi_+ (q), \\
  \LR V_2(g)\kern.2em\varphi_- \RR(p)
    &:=& e^{-ipa}\ D^{\KIn{$(\frac{n}{2},0)$}}(A)\ \varphi_- (q),
\end{eqnarray}
where $g=(A,a)\in\UPgr=\semi{\g{SL}{2,{\z{C}}}}{\z{R}^{4}}$ and 
$q:=\Lambda_{A}^{-1}p\in\Lkf$. Since 
\[
 \beta_+(q)=D^{\KIn{$(0,\frac{n}{2})$}}(A)^*\,
            \beta_+(p)\,D^{\KIn{$(0,\frac{n}{2})$}}(A)
 \quad\Wort{and}\quad
  \beta_-(q)=D^{\KIn{$(\frac{n}{2},0)$}}(A)^*\,
            \beta_-(p)\,D^{\KIn{$(\frac{n}{2},0)$}}(A)\,,
\]
for $p,q$ as before we have that the representations $V_{1,2}$ leave the 
sesquilinear forms $\langle\cdot,\cdot\rangle_{\beta_{\pm}}$ invariant. 
From the comments made at the beginning of this section about the
eigenvalues of $P^{\KIn{$\dagger$}}$ it is clear that the sesquilinear 
forms $\langle\cdot,\cdot\rangle_{\beta_{\pm}}$ are only semidefinite. 
This observation is in agreement with the general theorem in 
\cite[p.~113]{Barut72}. We can thus select in a natural way the 
following subspaces of ${\got H}_{n,\pm}$:

\begin{defi}
\label{Def.5.3.1}
With respect to the sesquilinear form defined above we can naturally
define:
\begin{eqnarray}
{\got H}_{n,+}^{(>)} &:=& 
    \LG \varphi\in{\got H}_{n,+}\mid  \varphi(p) =
    \mathop{\otimes}\limits^{n} 
            \left(\kern-1.5mm\begin{array}{c} 
                   -p_{1}-ip_{2} \\
                      p_0+p_3 
           \end{array}\kern-1.5mm\right) \chi_+(p),  
    \,\Wort{for suitable scalar}\; \chi_+ \RG                     \\[2mm]
{\got H}_{n,-}^{(>)} &:=& 
    \LG \varphi\in{\got H}_{n,-}\mid  \varphi(p) =
    \mathop{\otimes}\limits^{n} 
            \left(\kern-1.5mm\begin{array}{c} 
                   -p_{1}+ip_{2} \\
                      p_0+p_3 
           \end{array}\kern-1.5mm\right) \chi_-(p) ,  
     \,\Wort{for suitable scalar}\; \chi_-\RG                    \\[2mm]
{\got H}_{n,\pm}^{(0)} &:=& 
    \LG \varphi\in{\got H}_{n,\pm}\mid  
    \langle\varphi ,\varphi\rangle_{\beta_\pm}=0 \RG      \\[2mm]
{\got H}_{n,\pm}'      &:=&
        {{\got H}_{n,\pm}}/{{\got H}_{n,\pm}^{(0)}}
\end{eqnarray}
\end{defi}

\begin{lem}
\label{Lem.5.3.2}
Using the preceding definitions we have that for $n>0$ 
\begin{itemize} 
 \item[{\rm(i)}] ${\got H}_{n,\pm}=
                {\got H}_{n,\pm}^{(>)}\;\oplus\;{\got H}_{n,\pm}^{(0)}$.
 \item[{\rm(ii)}] The representations $V_{1,2}$ leave the spaces 
             ${\got H}_{n,\pm}^{(0)}$ invariant. On the contrary, the 
                         subspaces ${\got H}_{n,\pm}^{(>)}$
                  are not invariant under 
             the mentioned representations.  
\item[{\rm(iii)}] For any non zero $\varphi\in {\got H}_{n,\pm}^{(>)}$ we  
             have $\|\varphi\|_{\beta_{\pm}}=\|V_{1,2}(g)
             \varphi\|_{\beta_{\pm}}>0$ 
             for all $g\in \UPgr$.
\end{itemize} 
\end{lem}
\begin{beweis}
Part (i) follows directly from the analysis of the eigenvalues of the 
matrix $P^{\KIn{$\dagger$}}$ given at the beginning of this section.
That ${\got H}_{n,\pm}^{(0)}$ are $V_{1,2}$--invariant subspaces and 
part (iii) are a consequence of the fact that the representations
$V_{1,2}$ leave the sesquilinear forms 
$\langle\cdot,\cdot\rangle_{\beta_{\pm}}$
invariant. To prove the rest of part (ii) note that 
e.g.~for $f\in\Test{\z{R}^{4},{\cal H}}$ we have
\[
 \varphi_+^{\KIn{$(>)$}}(p):=
 D^{\KIn{$(0,\frac{n}{2})$}}\!\LR\!
         H_p \KIn{$\left(\kern-2mm\begin{array}{cc}
               0 \kern-2.6mm&  0\\ 
               0 \kern-2.6mm & 1
       \end{array} \kern-2mm\right)$}\!\RR \widehat{f}(p)
  \in{\got H}_{n,+}^{(>)}
     \quad\Wort{and}\quad
  \varphi_+^{\KIn{$(0)$}}(p):=
  D^{\KIn{$(0,\frac{n}{2})$}}\!\LR\!
         H_p \KIn{$\left(\kern-2mm\begin{array}{cc}
               1 \kern-2.6mm&  0\\ 
               0 \kern-2.6mm & 0
       \end{array} \kern-2mm\right)$}\!\RR \widehat{f}(p)
  \in{\got H}_{n,+}^{(0)}\,.
\]
Thus for a general $g=(A,0)\in\UPgr$ and since $H_p^{-1} A H_q\in
{\cal E}(2)$, $q:=\Lambda_{A}^{-1}p$,
\[
 (V_1(g)\varphi_+^{\KIn{$(>)$}})(p)
  = D^{\KIn{$(0,\frac{n}{2})$}}\!\LR\!
         A H_q \KIn{$\left(\kern-2mm\begin{array}{cc}
               0 \kern-2.6mm&  0\\ 
               0 \kern-2.6mm & 1
       \end{array} \kern-2mm\right)$}\!\RR \!\widehat{f}(q)
  = D^{\KIn{$(0,\frac{n}{2})$}}(H_p)\,
    D^{\KIn{$(0,\frac{n}{2})$}}\!\LR\!
         H_p^{-1} A H_q \KIn{$\left(\kern-2mm\begin{array}{cc}
               0 \kern-2.6mm&  0\\ 
               0 \kern-2.6mm & 1
       \end{array} \kern-2mm\right)$}\!\RR\! \widehat{f}(q)
  = \psi_+^{\KIn{$(>)$}}+\psi_+^{\KIn{$(0)$}} ,
\]
where $\psi_+^{(>)}\in {\got H}_{n,+}^{(>)}$, 
$\psi_+^{(0)}\in {\got H}_{n,+}^{(0)}$ (similar arguments for
the spaces with opposite helicity indexed with a `$-$'). 
This implies that the representations
$V_{1,2}$ restricted to ${\got H}_{n,\pm}^{(>)}$ produce in general 
further `zero norm vectors'.
\end{beweis}

From the preceding lemma we can lift the representations $V$ to the 
factor spaces ${\got H}_{n,\pm}'$. We denote the lift by $V'$
and the equivalence classes in ${\got H}_{n,\pm}'$ by $[\cdot]_\pm$.

\begin{teo}
\label{Teo.5.3.3}
The \rep{s} $V_{1,2}'$ defined on ${\got H}_{n,\pm}'$ are equivalent to
the irreducible and unitary Wigner \rep{s} $U_{\pm}$ defined in 
Eq.~$(\ref{Vf0Can})$.
\end{teo}
\begin{beweis}
We will give the proof for the spaces with index `+'. For the spaces with
opposite helicity similar arguments can be used just interchanging 
$D^{\KIn{$(0,\frac{n}{2})$}}(\cdot)$ with 
$D^{\KIn{$(\frac{n}{2},0)$}}(\cdot)$. For
$\chi\in\mathrm{L}^2(\Lkf,\z{C},\mudp)$ the linear mapping given by
\[
 (\Phi_+\,\chi)(p):= \left[ D^{\KIn{$(0,\frac{n}{2})$}}(H_p)
                     \LR \mathop{\otimes}\limits^n
                     \left(\kern-1.5mm  \begin{array}{c}  
                           0 \\ 1 
                     \end{array} \kern-1.5mm \right)
                     \RR\chi(p)\right]_+
\]
is easily seen to be an isometry between $\mathrm{L}^2(\Lkf,\z{C},\mudp)$
and ${\got H}_{n,+}'$ with the corresponding scalar products. Note that
$D^{\KIn{$(0,\frac{n}{2})$}}(H_p)
                     \LR \mathop{\otimes}\limits^n
                     \left(\kern-1.5mm  \begin{array}{c}  
                           0 \\ 1 
                     \end{array} \kern-1.5mm \right)
                     \RR\chi(p)$
is the representant in ${\got H}_{n,+}^{(>)}$ of the equivalence class.
Now the intertwining equation $\Phi_+\, U_+(g)=V_1(g)\,\Phi_+$, 
$g\in\UPgr$, is also a straightforward calculation if one recalls that
\[
 e^{\frac{i}{2}\theta(A,p)}= \left(\kern-1.5mm\begin{array}{cc}
                                0 \kern-1.6mm & 0 \\ 
                                0 \kern-1.6mm & 1
                                \end{array} \kern-1.5mm\right)
                                \overline{H_p^{-1}\,A\,H_q}
                                \left(\kern-1.5mm\begin{array}{cc}
                                0 \kern-1.6mm & 0 \\ 
                                0 \kern-1.6mm & 1
                                \end{array} \kern-1.5mm\right)
\quad\Wort{and that}\quad
 D^{\KIn{$(0,\frac{n}{2})$}}\LR H_p \left(\kern-1.5mm\begin{array}{cc}
                                  1 \kern-1.6mm & 0 \\ 
                                  0 \kern-1.6mm & 0
                                  \end{array} \kern-1.5mm\right)
                             \RR\varphi(p)\in{\got H}_{n,+}^{(0)}\,,
\]
for any suitable $\cal H$--valued function $\varphi$.
\end{beweis}
\begin{rem}
The representations $V_{1,2}$ on the spaces ${\got H}_{n,\pm}$
are the analogue of the massive representations 
that avoid the use of so--called `Wigner rotations'
(see e.g.~\cite{Niederer74a,Niederer74b} or \cite[Part~A]{Lledo95}).
The price for the masslessness here are the degenerate subspaces 
${\got H}_{n,\pm}^{(0)}$. The advantage of using these spaces is the fact
that one can more naturally reduce the covariant representation 
(\ref{VfCov}) in terms of $V_{1,2}'$ than in terms of $U_\pm$ 
given in eq.~(\ref{Vf0Can}). This reduction is done by means of the
embedding $\ot I.$, which is the
essential essential object for the construction of the free net
(see also \cite{Langbein67}).
\end{rem}

Recall next the covariant and massless canonical representations
(the latter written in the more convenient factor space notation)
considered before:
\[
 \LR T(g)\kern.2emf \RR\! (x)
   := D^{\KIn{$(0,\frac{n}{2})$}}(A)\, f\!\left( \Lambda_{A}^{-1}(x-a) \right),
  \quad g=(A,a)\in\UPgr\,,\;
  f\in\Test{\R^{4},{\cal H}^{\KIn{$(0,\frac{n}{2})$}}}\,,\\[-.35cm]
\]
\begin{eqnarray*}
\LR V_1'(g)\kern.2em[\varphi]_+\RR\!(p)
   \kern-3mm &:=&\kern-3mm 
    \LE e^{-ipa} D^{\KIn{$(0,\frac{n}{2})$}}(A)\varphi
    \!\left( \Lambda_{A}^{-1}p \right)\RE_+,
 \LR V_3'(g)\kern.2em[\varphi]_+ \RR\!(p)
     := \LE e^{ipa} D^{\KIn{$(0,\frac{n}{2})$}}(A)
           \varphi\!\left(\Lambda_{A}^{-1}p \right)\RE_+,\\
\LR V_2'(g)\kern.2em[\psi]_- \RR\!(p)
  \kern-3mm &:=&\kern-3mm 
     \LE e^{-ipa} D^{\KIn{$(\frac{n}{2},0)$}}(A)\psi
     \!\left(\Lambda_{A}^{-1}p\right)\RE_-,
 \LR V_4'(g)\kern.2em[\psi]_- \RR\!(p)
     := \LE e^{ipa}D^{\KIn{$(\frac{n}{2},0)$}}(A)\psi\!\left( 
           \Lambda_{A}^{-1}p \right)\RE_-\,,
\end{eqnarray*}
where $\varphi\in {\got H}_{n,+}$ and $\psi\in {\got H}_{n,-}$.
$V_1'$ and $V_2'$ satisfy the spectrality condition.

The following definition will be the essential ingredient for the massless
free net construction:
\begin{defi}\label{factornets}
Reference spaces and embeddings for the bosonic and fermionic cases:
\begin{itemize}
\item[{\rm (i)}] In the Bose case $(n$ even$)$
take $h_n:={\got H}_{n,+}'\oplus {\got H}_{n,-}'$ (considered as 
a real space) and the symplectic form
$\sigma_n:=\Wort{Im}\,\langle\cdot,\cdot\rangle_{\beta_+}
\oplus \Wort{Im}\,\langle\cdot,\cdot\rangle_{\beta_-}$. As symplectic 
\rep{} of $\UPgr$ choose $V_n:=V_1'\oplus V_2'$.
The embedding ${\got I}_n\colon
\Test{\R^{4},{\cal H}^{\KIn{$(0,\frac{n}{2})$}}}\to h_n$ is given here by
\[
 ({\got I}_n f)(p):=[\widehat{f}(p)]_+\oplus [\widehat{\Gamma_0 f}(p)]_-\,,
 p\in\Lkf\,,
\]
where $\Gamma_0\colon\ {\cal H}^{\KIn{$(0,\frac{n}{2})$}}\to
{\cal H}^{\KIn{$(\frac{n}{2},0)$}}$ is an antiunitary involution
and $\widehat{f}(p):=\int_{\R^4}e^{-ipx}f(x)\,\mathrm{d}^4x$ is the 
Fourier transform.

\item[{\rm (ii)}] In the Fermi case $(n$ odd$)$
take $h_n:={\got H}_{n,+}'\oplus {\got H}_{n,-}'\oplus {\got H}_{n,+}'
\oplus {\got H}_{n,-}'$ with the natural scalar product and the antilinear
involution given by
\[
 \Gamma_n\LR[\varphi_+]_+\oplus[\varphi_-]_-
         \oplus[\psi_+]_+\oplus[\psi_-]_-\RR:=
 [\Gamma_0\,\psi_-]_+\oplus [\Gamma_0\,\psi_+]_-\oplus
  [\Gamma_0\,\varphi_-]_+\oplus[\Gamma_0\,\varphi_+]_-
\]
As unitary \rep{} of $\UPgr$ that intertwines with $\Gamma_n$
choose $V_n:=V_1'\oplus V_2'\oplus V_3'\oplus V_4'$.
In the present case the embedding 
${\got I}_n\colon
\Test{\R^{4},{\cal H}^{\KIn{$(0,\frac{n}{2})$}}}\to h_n$ is given by
\[
 ({\got I}_n f)(p):=[\widehat{f}(p)]_+\oplus [\widehat{\Gamma_0 f}(p)]_-
   \oplus [\widehat{f}(-p)]_+\oplus [\widehat{\Gamma_0 f}(-p)]_-\,,
 p\in\Lkf\,.
\]
\end{itemize}
\end{defi}

The preceding embeddings characterize in a canonical way nets of
C*--subalgebras of the CAR-- and CCR--algebras associated to
the corresponding reference space $h_n$ (cf.~\cite[Chapter~8]{bBaumgaertel92} 
and references cited therein). The explicit construction of the net
and the verification of some of the main axioms of algebraic 
QFT is the content of the following theorem.

\begin{teo}\label{Prop}
Denoting by $\BR$ the set of open and bounded regions in Minkowski space we
have  
\begin{itemize}
\item[{\rm (i)}] Fermionic case ($n$ odd):
\[
 \BR\ni {\cal O}\longmapsto {\cal A}_n({\cal O})
 :=\mathrm{C}^*\LR \{A(\ot I._nf)\mid\mathrm{supp}f \subset {\cal O} \}
 \RR^{\KIn{$\z{Z}_2$}} \subset \car{h_n,\Gamma_n}\,.
\]
Here $A(\cdot)$ are the generators of the CAR--algebra
$\car{h_n,\Gamma_n}$ and
${\cal A}^{\;\KIn{$\z{Z}_2$}}$ denotes the fixed point subalgebra of the 
C*--algebra $\cal A$ w.r.t.~Bogoljubov automorphism associated to the 
unitarity $-\EINS$. 
\item[{\rm (ii)}] Bosonic case ($n$ even):
\[
 \BR\ni {\cal O}\longmapsto {\cal A}_n({\cal O}) 
 :=\mathrm{C}^*\!\LR \{\delta_{\ot I._nf}\mid\mathrm{supp}f 
   \subset {\cal O} \}\RR\subset\ccr{h_n,\sigma_n}\,.
\]
Here $\delta_{(\cdot)}$ denote the Weyl elements that generate the CCR--algebra
$\ccr{h_n,\sigma_n}$.
\end{itemize}
Finally, the net $\BR\ni {\cal O}\mapsto {\cal A}_n({\cal O})$ characterized 
by the corresponding embeddings ${\got I}_n$, $n\in\N$,
satisfies the properties of 
\begin{itemize}
\item[{\rm (i)}] $($Isotony$)$
 If $\B_1\subseteq\B_2$, then
 $\al A._n(\B_1)\subseteq\al A._n(\B_2)$, $\al O._1,\al O._2\in\BR$.
\item[{\rm (ii)}] $($Causality$)$
 If $\B_1$ and $\B_2$ are causaly separated, 
 then $[\al A._n(\B_1),\al A._n(\B_2)]=0$.
\item[{\rm (iii)}] $($Additivity$)$
 For any $\{\al O._\lambda\}_{\lambda\in\Lambda}\subset\BR$ with
 $\cup_\lambda\al O._\lambda\in\BR$. Then 
\[
 \al A._n(\cup_\lambda\al O._\lambda)= 
   \mathrm{C}^*\Big(\cup_\lambda\al A._n(\al O._\lambda)\Big)\,.
\] 
\item[{\rm (iv)}] $($Covariance$)$ 
 There exists a representation 
 $\UPgr\ni g\mapsto\alpha_g$ in terms of automorphisms
 of the CAR--resp.~CCR--algebras such that $\alpha_g (\al A._n(\al O.))=
 \al A.(g\al O.)$, $g\in\UPgr$, $\al O.\in\BR$.
\end{itemize}
\end{teo}
\begin{beweis}
Since in this paper the covariance axiom plays a distinguished 
role we will show only this property here. For the other properties 
and further details we refer to \cite{Lledo95,tLledo98,pLledo99b}. 
We will show that the covariance relation
is based on the following intertwining property of the 
embeddings $\ot I._n$ w.r.t.~the covariant and the canonical
representations: 
\[
 \ot I._n T(g)=V_n(g) \ot I._n\,,\quad g\in\UPgr\,.
\]
Indeed, let $\alpha_g$ be the Bogoljubov automorphisms associated to the
Bogoljubov unitaries $V_n(g)$. Further note also that the covariant 
representation $T$ shifts the space time regions in the correct way,
i.e.~if $f\in\Test{\R^{4},{\cal H}}$ and $\mathrm{supp}f\subset\al O.$,
then $\mathrm{supp}(T(g)f)\subset g\al O.$, $g\in\UPgr$. Now in the 
bosonic case ($n$ even) we have for any $\al O.\in\BR$, $g\in\UPgr$,
\begin{eqnarray*}
  \alpha_g\Big(\al A.(\al O.)\Big)
     &=& \alpha_g\Big(\mathrm{C}^*\!\LR \{\delta_{\ot I._nf}\mid\mathrm{supp}f 
                                        \subset {\cal O} \}\RR\Big)\\
     &=& \mathrm{C}^*\!\LR \{\alpha_g(\delta_{\ot I._nf})\mid\mathrm{supp}f 
         \subset {\cal O} \}\RR \\
     &=& \mathrm{C}^*\!\LR \{\delta_{V_n(g)(\ot I._nf)}\mid\mathrm{supp}f 
                                        \subset {\cal O} \}\RR\\
     &=& \mathrm{C}^*\!\LR \{\delta_{\ot I._n(T(g)f)}\mid\mathrm{supp}f 
                                        \subset {\cal O} \}\RR\\
     &=& \mathrm{C}^*\!\LR \{\delta_{\ot I._nf'}\mid\mathrm{supp}f'
                                        \subset g{\cal O} \}\RR\\
     &=& \al A.(g\al O.)\,.
\end{eqnarray*}
One can argue similarly for the fermionic nets.
\end{beweis}

\begin{rem}\label{OneP}
Note that for the free nets constructed previously the one particle 
Hilbert space corresponding to the canonical Fock states (to be specified
in Section~\ref{conseq}) is ${\got H}_{n,+}'\oplus {\got H}_{n,-}'$, $n\in\N$.
It carries a representation $V_1'\oplus V_2'$ which by 
Theorem~\ref{Teo.5.3.3} is equivalent to the reducible Wigner massless
representations $U_+\oplus U_-$ with helicities $\frac{n}{2}$ and
$-\frac{n}{2}$. This Hilbert space and representation coincide with the
one--particle Hilbert space used in more standard quantum field theoretical
construction of massless free fields (cf.~e.g.~\cite{Hislop88}).

We will show later that the covariance property of the
massless free nets can be extended to the 
fourfold coverning of the conformal group $\g{SU}{2,2}$.
\end{rem}

\section{Extension of the massless representations}

The first step to show that the massless free nets contructed in the
previous theorem are also covariant w.r.t.~the conformal group is to
show that the massless canonical representations $V_k'$, $k=1,2,3,4$,
extend within $\ot H._\pm'$ to a unitary representation of 
$\g{SU}{2,2}$. This fact has been shown considering different mathematical
contexts (see e.g.~\cite{Mack69,Carey77c,Jakobsen77,Angelopoulos81}). We will
give next a simplified proof of this result due to the nice properties
the functions $\beta_\pm(\cdot)$ introduced in the previous section.

In the context of Subsection~\ref{subconf} we consider as 
inducing \rep{} of the little group ${\cal K}_0$ on
$\al H.^{\KIn{$(0,\frac{n}{2})$}}$
\[
 \tau(K):=\Wort{det}(A-iB)\,D^{\KIn{$(0,\frac{n}{2})$}}(A-iB)\,,\quad
          K=\left(\kern-1.5mm \begin{array}{cc}
             A \kern-2mm &  B \\
            -B \kern-2mm &  A 
        \end{array} \kern-1.5mm\right)\in{\cal K}_0 \,,
\]
which is unitary because from Eqs.~(\ref{ABCD}) we have in this case
$AA^*+BB^*=\EINS$, $AB^*=-BA^*$ etc.
The first step will be to define a mapping $J$
that satisfies the properties required in Remark~\ref{HFunct}~(ii).

\begin{lem}\label{LDefJ} 
For $g=\left(\kern-1.5mm \begin{array}{cc}
                      A_1 \kern-2mm &  A_2  \\
                      A_3 \kern-2mm &  A_4
     \end{array} \kern-1.5mm\right)\in\g{SU}{2,2}$, $Z\in\TT$,
the mapping
\[
 J^{\KIn{$(0,\frac{n}{2})$}}(g,Z):=
  \Wort{det}(A_3Z+A_4)\,D^{\KIn{$(0,\frac{n}{2})$}}(A_3Z+A_4)
\] 
satisfies the properties 
\begin{eqnarray*}
  J^{\KIn{$(0,\frac{n}{2})$}}(g_1g_2,Z) 
         &=& J^{\KIn{$(0,\frac{n}{2})$}}(g_1,g_2 Z)\,
             J^{\KIn{$(0,\frac{n}{2})$}}(g_2,Z)\,,
             \quad g_1,g_2\in\g{SU}{2,2}\,,\;Z\in\TT \,,\\
 J^{\KIn{$(0,\frac{n}{2})$}}(\EINS,Z)  &=&\EINS \,, \\ 
 J^{\KIn{$(0,\frac{n}{2})$}}(K,i\EINS) &=& \tau(K) \,,\quad K\in\al K._0\,.
 \end{eqnarray*}
\end{lem}
\begin{beweis}
The second property is trivial since $J^{\KIn{$(0,\frac{n}{2})$}}(\EINS,Z)
=\Wort{det}(\EINS)\,D^{\KIn{$(0,\frac{n}{2})$}}(\EINS)=\EINS_{\al H.}$.
Further, the third eq.~follows also directly from the choices of $\tau$
and $J$. To prove the
first property put $J_1(g,Z):=A_3Z+A_4$ (which acts on $\C^2$)
and note that it already satisfies the condition
$J_1(g_1g_2,Z) = J_1(g_1,g_2 Z)\,J_1(g_2,Z)$ as can be immediately 
checked. Therefore, since $D(\cdot)$ is a \rep{} and from the product
rule for determinants it follows that 
$J(g,Z)=\Wort{det}(J_1(g,Z))\,D(J_1(g,Z))$ satisfies the required condition. 
\end{beweis}

\begin{rem}\label{better}
The following results will be close to those in
\cite[Section~IV]{Jakobsen77}. The main difference w.r.t.~Jakobsen
and Vergne's 
approach lies in the fact the we are working with Minkowski scalar
products in the arguments of the exponentials that appear, while
in the cited reference mainly euclidean scalar products are considered.
This variation will have no consequence for the absolute convergence 
of the integrals studied next and it will considerably simplify some
proofs later on, e.g.~the extension result for the massless canonical 
\rep{} of the Poincar\'e group (cf.~Theorem~\ref{ExtW1}). 
Note that we can write $xp=x_0p_0-\sum_i x_ip_i=\Wort{Tr}(P^\dagger X)$ and 
$x_0p_0+\sum_i x_ip_i=\In{$\frac12$}\Wort{Tr}(PX)$, where $P,P^\dagger$
are given in Section~\ref{mfn}. 
The following two technical lemmas will be essential for the proof of the
conformal covariance of massless free nets.
\end{rem}

\begin{lem}\label{LAC1}
For $Y\in\g{H$_+$}{2,\C}$ we have
\[
 \int\limits_{\Lkf}e^{-\KIn{Tr$(P^\dagger Y)$}}\beta_+(p) \mudp
  = C_n\,(\Wort{det}\,Y)^{-1}\,D^{\KIn{$(0,\frac{n}{2})$}}(Y)^{-1}\,,\quad
    C_n>0\,,
\]
where the l.h.s.~is an absolutely convergent integral.
\end{lem}
\begin{beweis}
First note that with the notation above 
$\overline{P^\dagger}=
    \frac12\KIn{$\left(\kern-2mm\begin{array}{cc}
               0 \kern-2.6mm& -1\\ 
               1 \kern-2.6mm & 0
       \end{array} \kern-2mm\right)$} P
            \KIn{$\left(\kern-2mm\begin{array}{cc}
               0 \kern-2.6mm & 1\\ 
              -1 \kern-2.6mm & 0
       \end{array} \kern-2mm\right)$}$ so that 
\bea
 \lefteqn{\nonumber\!\!\!\!\!\!\!\!\!\!\!
  \int\limits_{\Lkf}e^{-\KIn{Tr$(P^\dagger Y)$}}\beta_+(p) \mudp}\\
  &=&\nonumber\LR\In{$\frac12$}\RR^n\, D^{\KIn{$(0,\frac{n}{2})$}}
     \KIn{$\LR\!\left(\kern-2mm\begin{array}{cc}
               0 \kern-2.6mm & -1\\ 
               1 \kern-2.6mm & 0
       \end{array} \kern-2mm\right)\!\RR$}         \\ 
  & &\label{AbsConv} \LR\;\int\limits_{\Lkf}
   e^{-\KIn{Tr$\LR P
   \left(\kern-2mm\begin{array}{cc}
               0 \kern-2.6mm & 1\\ 
               -1 \kern-2.6mm & 0
       \end{array} \kern-2mm\right)
    (\frac12\,Y^t)
   \left(\kern-2mm\begin{array}{cc}
               0 \kern-2.6mm & -1\\ 
               1 \kern-2.6mm & 0
   \end{array} \kern-2mm\right)\!
  \RR$}} D^{\KIn{$(\frac{n}{2},0)$}}(P) \mudp\RR
  D^{\KIn{$(0,\frac{n}{2})$}}
       \KIn{$\LR\!\left(\kern-2mm\begin{array}{cc}
               0 \kern-2.6mm & 1\\ 
               -1 \kern-2.6mm & 0
       \end{array} \kern-2mm\right)\!\RR$} ,
\eea
where the index $t$ means matrix transposition. But from 
\cite[Proposition~IV.1.1]{Jakobsen77} the integral on the r.h.s.~of the
preceding equation is absolutely convergent for $Y\in\g{H$_+$}{2,\C}$ 
and even more we also have from the mentioned proposition
that for some $C_n'>0$ 
\[
 \int\limits_{\Lkf}
   e^{-\KIn{Tr$\LR\! P
   \left(\kern-2mm\begin{array}{cc}
               0 \kern-2.6mm & 1\\ 
               -1 \kern-2.6mm & 0
       \end{array} \kern-2mm\right)
     (\frac12\,Y^t)
   \left(\kern-2mm\begin{array}{cc}
               0 \kern-2.6mm & -1\\ 
               1 \kern-2.6mm & 0
   \end{array} \kern-2mm\right)\!
  \RR$}} D^{\KIn{$(\frac{n}{2},0)$}}(P) \mudp
 = C_n'\,(\Wort{det}\,Y)^{-1}\,
   D^{\KIn{$(\frac{n}{2},0)$}}
   \LR
\In{$\frac12$}\KIn{$\left(\kern-2mm\begin{array}{cc}
               0 \kern-2.6mm&  1\\ 
               -1 \kern-2.6mm & 0
       \end{array} \kern-2mm\right)$} \,Y^t\,
            \KIn{$\left(\kern-2mm\begin{array}{cc}
               0 \kern-2.6mm & -1\\ 
               1 \kern-2.6mm & 0
       \end{array} \kern-2mm\right)$}
   \RR^{-1} .
\]
Inserting this result on the r.h.s.~of Eq.~(\ref{AbsConv})
and using $D^{\KIn{$(\frac{n}{2},0)$}}(Y^t)=
D^{\KIn{$(0,\frac{n}{2})$}}(Y)$ we get the eq.~of the lemma.
\end{beweis}

\begin{lem}\label{LAC2}
For $Z_1,Z_2\in\TT$ we have 
\[
 {\got K}_+(Z_1,Z_2)
    :=C_n\LR\Wort{det}\LR
      \frac{\scriptstyle Z_1-Z_2^*}{\scriptstyle 2i}\RR\!\RR^{-1}
      D^{\KIn{$(0,\frac{n}{2})$}}\LR
      \frac{\scriptstyle Z_1-Z_2^*}{\scriptstyle 2i}\RR^{-1}
     = \int\limits_{\Lkf}
       e^{i\,\KIn{Tr$\left(P^\dagger(Z_1-Z_2^*)\right)$}}\beta_+(p)\mudp\,,
\]
where the integral is absolutely convergent. Further, for any $g\in
\g{SU}{2,2}$ we have
\[
 {\got K}_+(gZ_1,gZ_2)=J^{\KIn{$(0,\frac{n}{2})$}}(g,Z_1)\,{\got K}_+(Z_1,Z_2)
                 \LR J^{\KIn{$(0,\frac{n}{2})$}}(g,Z_2)\RR^*
\]
\end{lem}
\begin{beweis}
Note first that if $Z_1,Z_2\in\TT$, then $Z_1-Z_2^*\in\TT$, which 
implies $\Wort{det}(Z_1-Z_2^*)\not= 0$. Applying now Lemma~\ref{LAC1}
as well as \cite[Proposition~IV.1.2]{Jakobsen77} we get the first part of
the statement.

To prove the last equation take $g=\left(\kern-1.5mm \begin{array}{cc}
                      A_1 \kern-2mm &  A_2  \\
                      A_3 \kern-2mm &  A_4
                 \end{array} \kern-1.5mm\right)\in\g{SU}{2,2}$
and consider first
\begin{eqnarray*}
 \frac{1}{2i}\LR gZ_1-(gZ_2)^*\RR^{-1}
 \kern-4mm&=&\kern-3mm 
     \frac{1}{2i}\LR
     (A_1Z_1+A_2)(A_3Z_1+A_4)^{-1}-((A_1Z_2+A_2)(A_3Z_2+A_4)^{-1})^* 
     \RR^{-1}                                                      \\[2mm]
 \kern-4mm&=&\kern-3mm \frac{1}{2i}\,(A_3Z_1+A_4)\cdot             \\
 \kern-4mm& &\kern-3mm 
     \LR(Z_2^*A_3^*\!+\!A_4^*)(A_1Z_1\!+\!A_2)-(Z_2^*A_1^*\!+\!A_2^*)
      (A_3Z_1\!+\!A_4)\RR^{-1}\!\cdot(A_3Z_2+A_4)^*                \\[2mm]
 \kern-4mm&=&\kern-3mm
     (A_3Z_1+A_4)\cdot\frac{1}{2i}(Z_1-Z_2^*)^{-1}\cdot(A_3Z_2+A_4)^*\,,
\end{eqnarray*}
where for the last eq.~we have used the relations (\ref{ABCD}). 
Now recalling the definition of $J$ in Lemma~\ref{LDefJ} 
we have that
\begin{eqnarray*}
{\got K}_+(gZ_1,gZ_2) 
 &=& C_n\,\LR\Wort{det}\LR
     \frac{\scriptstyle gZ_1-(gZ_2)^*}{\scriptstyle 2i}\RR\!\RR^{-1}
     \,D^{\KIn{$(0,\frac{n}{2})$}}\LR
     \frac{\scriptstyle gZ_1-(gZ_2)^*}{\scriptstyle 2i}\RR^{-1}\\
 &=& J^{\KIn{$(0,\frac{n}{2})$}}(g,Z_1)\,{\got K}_+(Z_1,Z_2)
       \LR J^{\KIn{$(0,\frac{n}{2})$}}(g,Z_2)\RR^*
\end{eqnarray*}
and the proof is concluded.
\end{beweis}

We will explicitly give next a parametrization in terms of the sets
$\TT$ and $\al H.$ of a total
set ${\got H}_{tot}'\subset{\got H}_{n,+}'$
that will satisfy the properties 
required in Subsection~\ref{subconf}.
\begin{lem}\label{KCov}
The set 
${\got H}_{tot}':=\{\, [K_{Z,v}]_+\mid K_{Z,v}(p):=
                  e^{-i\KIn{Tr$(P^\dagger Z^*)$}}v\,,\;
                  p\in\Lkf\,,Z\in\TT\,,v\in\al H.\}$
is total in ${\got H}_{n,+}'$. Further the following equation holds
for all $g\in\g{SU}{2,2}$:
\[
 \askp{K_{gZ_1,v_1}}{K_{gZ_2,v_2}}_{\beta_+}
  =\askp{K_{Z_1,\,J(g,Z_1)^*v_1}}{K_{Z_2,\,J(g,Z_2)^*v_2}}_{\beta_+}
   \,,\quad Z_1,Z_2\in\TT\,,\;v_1,v_2\in\al H.\,.
\]
\end{lem}
\begin{beweis}
First from Lemma~\ref{LAC1} we have for any $Z=X+iY\in\TT$, 
$v\in\al H.$, that
\[
 \askp{K_{Z,v}}{K_{Z,v}}_{\beta_+}=
   \int\limits_{\Lkf} e^{-\KIn{Tr$(2P^\dagger Y)$}}
                      \askp{v}{\beta_+(p)v}_{\al H.}\mudp < \infty
\]
and $\Wort{span}\,{\got H}_{tot}'$ is dense in ${\got H}_{n,+}'$ by 
Lemmas~4.2.2 and 4.2.3 in \cite{Carey77c}.

Finally, recalling the properties of ${\got K}_+$ in Lemma~\ref{LAC2} we have
\begin{eqnarray*}
\askp{K_{gZ_1,v_1}}{K_{gZ_2,v_2}}_{\beta_+}
 \kern-3mm &=&\kern-3mm \int\limits_{\Lkf} 
     \askp{e^{-i\KIn{Tr$(P^\dagger (gZ_1)^*)$}}\,v_1}{\beta_+(p)\,
     e^{-i\KIn{Tr$(P^\dagger (gZ_2)^*)$}}\,v_2}_{\al H.}\mudp      \\[2mm]
 \kern-3mm &=&\kern-3mm \bigg\langle v_1\,,\,
   \underbrace{\left(\;\int\limits_{\Lkf}
   e^{i\,\KIn{Tr$\left( P^\dagger(gZ_1-(gZ_2)^*)\right)$}}
    \beta_+(p) \mudp\right)}_{\scriptstyle {\got K}_+(gZ_1,gZ_2) }
     v_2\bigg\rangle_{\al H.}                                           \\
  \kern-3mm &=&\kern-3mm \int\limits_{\Lkf} 
    \askp{e^{-i\KIn{Tr$(P^\dagger Z_1^*)$}}
    \,J(g,Z_1)^*v_1}{\beta_+(p)\,
    e^{-i\KIn{Tr$(P^\dagger Z_2^*)$}}
     \,J(g,Z_2)^*v_2}_{\al H.}\mudp \\[2mm]
   \kern-3mm &=&\kern-3mm
   \askp{K_{Z_1,\,J(g,Z_1)^*v_1}}{K_{Z_2,\,J(g,Z_2)^*v_2}}_{\beta_+}\,,
\end{eqnarray*}
and the proof is concluded.
\end{beweis}

\begin{teo}\label{ExtW1}
The following \rep{} of the conformal group defined on ${\got H}_{tot}'$ by
\[
 W_1'(g)\, [K_{Z,v}]_+(p):=\LE K_{gZ,\,(J(g,Z)^{-1})^*v}\RE_+(p)
 \,,\quad g\in\g{SU}{2,2}\,,
\]
extends to a unitary and irreducible \rep{} within ${\got H}_{n,+}'$.
Further, the restriction of $W_1'$ to the Poincar\'e subgroup coincides
with $V_1'$ defined in Section~$\ref{mfn}$, which is equivalent to 
the massless canonical \rep{} of helicity $\frac{n}{2}$.
\end{teo}
\begin{beweis}
First of all note that by the proof of Lemma~\ref{KCov} we have that
$K_{Z,v}\mapsto K_{gZ,\,(J(g,Z)^{-1})^*v}$ leaves the sesquilinear 
form $\langle\cdot,\cdot\rangle_{\beta_+}$ invariant and therefore
the definition of $W_1'$ on the factor space is consistent with the
corresponding equivalence classes. Now again by Lemma~\ref{KCov}
we can apply Lemma~\ref{TConfUnit} to the present situation 
to conclude that $W_1'$ extends to a unitary \rep{} within ${\got H}_{n,+}'$.

Consider next the Poincar\'e subgroup of $\g{SU}{2,2}$ given in 
Eq.~(\ref{SubPoin}), i.e.
\[
 \g{SU}{2,2}\supset \UPgr\ni g_0=
  \left(\kern-1.5mm \begin{array}{cc}
                      A \kern-2mm &  C (A^*)^{-1} \\
                      0 \kern-2mm &  (A^*)^{-1}
                 \end{array} \kern-1.5mm\right)\,, A\in\g{SL}{2,\C}\,,
           C=C^* \in \g{H}{2,\C}\,.
\]
For this subgroup and recalling the action in Eq.~(\ref{CAction}) as well 
as Remark~\ref{better} we have
\[
 e^{-i\KIn{Tr$(P^\dagger\, (g_0Z)^*)$}}
  = e^{-i\KIn{Tr$(P^\dagger C)$}}\;
    e^{-i\KIn{Tr$\left(P^\dagger\, AZ^*A^*\right)$}}
  = e^{-i\KIn{Tr$(P^\dagger C)$}}\;
    e^{-i\KIn{Tr$\left( (A^{-1}P(A^{-1})^*)^\dagger\,Z^*\right)$}}
\]
and also $J^{\KIn{$(0,\frac{n}{2})$}}(g_0,Z)=D^{\KIn{$(0,\frac{n}{2})$}}
(A^{-1})^*$. From this we get
\[
 W_1'(g_0)\, [K_{Z,v}]_+(p)=
   \LE e^{-i\KIn{Tr$(P^\dagger C)$}}\,
D^{\KIn{$(0,\frac{n}{2})$}}(A)\,K_{Z,v}\left(\Lambda_{A}^{-1}p \right)\RE_+ .
\]
Thus the unitary \rep{s} $W_1'\hrist\UPgr$ and $V_1'$ coincide
on a total set and therefore they must be equal. Now, $V_1'$
is equivalent to massless canonical \rep{} with helicity 
$\frac{n}{2}$ and since $V_1'=W_1'\hrist\UPgr$ is already
irreducible, then $W_1'$ is certainly irreducible for the whole
$\g{SU}{2,2}$.
\end{beweis}

\begin{rem}\label{OtherSpaces}
\begin{itemize}
\item[(i)] Note that the representation $W_1'$ is just the transcription
  of the induced representation considered Remark~\ref{HFunct}~(ii)
  in terms of the more useful set of functions ${\got H}_{tot}'$.
  Indeed, recalling the kernels introduced in Lemma~\ref{LAC2} consider
  the following functions $\varphi_{\KIn{$Z_0$},v}\colon\ 
  \TT\to\al H.$, $Z_0\in\TT$, $v\in\al H.$,
 \[
   \varphi_{\KIn{$Z_0$},v}(Z):={\got K}_+(Z_0,Z)\,v
       = \int\limits_{\Lkf} e^{i\,\KIn{Tr$\left(P^\dagger Z_0\right)$}}
         \beta_+(p)\,K_{Z,v}(p) \,\mudp \;.
 \]
Using again Lemma~\ref{LAC2} it is now straightforward to rewrite the
induced representation (\ref{KerJ}) for the functions 
$\varphi_{\KIn{$Z_0$},v}$ in terms of the functions $K_{Z,v}$.
\item[(ii)] We can argue similarly as in this section for the spaces
with opposite helicity. Indeed, use the mapping 
$J^{\KIn{$(\frac{n}{2},0)$}}(g,Z):=\Gamma_0\,
J^{\KIn{$(0,\frac{n}{2})$}}(g,Z)\,\Gamma_0$  
and the kernel ${\got K}_-(Z_1,Z_2)=
\Gamma_0\,{\got K}_+(Z_1,Z_2)\,\Gamma_0$. 
It can be easily seen now that we can
extend as in the preceding theorem the \rep{s}
$V_i'$ needed in the previous section to the define the
free nets to corresponding representations $W_i'$, $i=2,3,4$.
\end{itemize}
\end{rem}


\section{Conformal covariance and its consequences}\label{conseq}

One of the characteristic facts about the conformal group is that 
it acts quasi--globaly on Minkowski space. This behaviour 
is due to the fact that the subgroup of the special conformal 
transformations has always singularities on certain hypersurfaces
of $\R^4$. We will therefore restrict in this section to 
$g\in\g{SU}{2,2}$, $f\in\TestO{\R^{4},{\cal H}}$
and double cones $\al O.\in\BR$, where $g\,\Wort{supp}f$ and
$g\al O.\subset \R^4$ are well defined.
We denote the family of double cones in $\R^4$ by $\al K.$.
These are standard assumptions in order to understand the axiom of covariance
in the general setting of conformal quantum field theory
(cf.~\cite[Section~1]{Brunetti93}, \cite[I.4]{Todorov78}).

To apply next the explicit formulas concerning the
\rep{s} of $\g{SU}{2,2}$ considered in the preceding section, we
will need to introduce first a suitable $y$--dependent embedding
($y\in\al V._+$, i.e. $Y\in\g{H$_+$}{2,\C}$) which can be related
to the embedding needed in Section~\ref{mfn} to the define the free net.
\begin{defi}\label{YEmbed}
Putting $Z=X+iY\in\TT$ we define ${\got I}_{y,+}\colon\
\TestO{\R^{4},{\cal H}^{\KIn{$(0,\frac{n}{2})$}}} \to 
{\got H}_{n,+}'$ by
\[ 
 \LR {\got I}_{y,+}f\RR\!(p):=\LE\; \int\limits_{\R^4}
  e^{-i\,\KIn{Tr$\left( P^\dagger Z^* \right)$}} f(x)\Wort{d}^4x \RE_+
                 =\LE\; \int\limits_{\R^4}
  e^{-i\,\KIn{Tr$\left( P^\dagger (X-iY) \right)$}} f(x)\Wort{d}^4x \RE_+ .
\]
\end{defi}

\begin{lem}\label{Limit}
The $y$--dependent embedding satisfies ${\got I}_{y,+}f
\in {\got H}_{n,+}'$ and          
$\lim\limits_{{\cal V}_+\ni y\to 0}{\got I}_{y,+}f=
[\,\widehat{f}\,]_+$, where the limit exists in the 
Hilbert space norm $\|\cdot\|_{\beta_+}$.
\end{lem}
\begin{beweis}
For any $f\in\TestO{\R^{4},{\cal H}^{\KIn{$(0,\frac{n}{2})$}}}$ it 
follows from Lemma~\ref{LAC1} that ${\got I}_{y,+}f\in {\got H}_{n,+}'$.
Further
\begin{eqnarray*}
\|{\got I}_{y,+}f-[\,\widehat{f}\,]_+\|_{\beta_+}^2
  &=& \int\limits_{\Lkf}
      \askp{\left({\got I}_{y,+}f(p)-\widehat{f}(p)\right)}{\beta_+(p)\,
      \left({\got I}_{y,+}f(p)-\widehat{f}(p)\right)}_{\al H.}\mudp  \\
  &=& \int\limits_{\R^4} \int\limits_{\R^4}\;\int\limits_{\Lkf}
      |e^{-\KIn{Tr$(P^\dagger Y)$}}-1 |^2 \;
      \askp{f(x)}{\beta_+(p)\,f(x')}\mudp\;\Wort{d}^4x \Wort{d}^4x'
\end{eqnarray*}
and the last expression tends to zero as ${\cal V}_+\ni y\to 0$
by Lebesgue's dominated convergence theorem 
(note that for $y\in {\cal V}_+$ we have
$1\geq |e^{-\KIn{Tr$(P^\dagger Y)$}}-1 |^2 \to 0$ as
${\cal V}_+\ni y\to 0$).
\end{beweis}

Now inspired by Theorem~\ref{ExtW1} we can consider the following 
\rep{} on the set of embedded test functions.

\begin{defi}
For $f\in\TestO{\R^{4},{\cal H}^{\KIn{$(0,\frac{n}{2})$}}}$,
$g\in\g{SU}{2,2}$ and $Y\in \g{H$_+$}{2,\C}$ we define
\[ 
 \LR W_1(g)({\got I}_{y,+}f)\RR\!(p)
   :=\LE\; \int\limits_{\R^4}
      e^{-i\,\KIn{Tr$\left( P^\dagger (gZ)^* \right)$}} 
      \left(J^{\KIn{$(0,\frac{n}{2})$}}(g,Z)^{-1}\right)^* 
      f(x)\Wort{d}^4x \RE_+ .
\]
\end{defi}

\begin{lem}
The \rep{} defined before satisfies for 
$f,k\in\TestO{\R^{4},{\cal H}^{\KIn{$(0,\frac{n}{2})$}}}$
\[
 \askp{W_1(g){\got I}_{y,+}f}{W_1(g){\got I}_{y,+}k}_{\beta_+}
  =\askp{{\got I}_{y,+}f}{{\got I}_{y,+}k}_{\beta_+}\,, 
    \quad g\in\g{SU}{2,2}\,, Z=X+iY\in\TT \,. 
\]
Further we have $W_1(g)\,({\got I}_{y,+}f)={\got I}_{gy,+}\,(T_y(g)f)$, where
\[
 \left(T_y(g)f\right)(gx)
    :=\left(J^{\KIn{$(0,\frac{n}{2})$}}(g,Z)^{-1}\right)^* f(x)
\]
satisfies the relation $T_y(g_1g_2)=T_{g_2y}(g_1)\,T_y(g_2)$,
$g_1,g_2\in\g{SU}{2,2}$.
\end{lem}
\begin{beweis}
The unitarity property is based on Lemma~\ref{LAC2}
(cf.~with the proof of Theorem~\ref{ExtW1}).
The other relations follow immediately from the definition
of $T_y$.
\end{beweis}

\begin{teo}\label{DefT0}
Consider for suitable 
$f\in\TestO{\R^{4},{\cal H}^{\KIn{$(0,\frac{n}{2})$}}}$ and
$g\in\g{SU}{2,2}$ the \rep{} 
\[
 \left(W_1(g) [\,\widehat{f}\,]_+\right)(p)
  := \lim\limits_{{\cal V}_+\ni gy\to 0}
     \LR W_1(g)({\got I}_{y,+}f)\RR\!(p)
   = \LE \left( \widehat{T_0(g)f}\right)\!(p)  \RE_+ ,
\]
where $\left( T_0(g)f \right)\!(gx):=
\left(J^{\KIn{$(0,\frac{n}{2})$}}(g,X)^{-1}\right)^* f(x)$.
Further $W_1\hrist\UPgr=V_1'$ on the set
$\{\,[\,\widehat{f}\,]_+\mid 
f\in\TestO{\R^{4},{\cal H}^{\KIn{$(0,\frac{n}{2})$}}}\}$ and
$T_0$ is a representation of $\g{SU}{2,2}$ on the test functions
that satisfies $\Wort{supp}(T_0(g)f)\subseteq g\,\Wort{supp}f$.
Finally, $T_0\hrist\UPgr$ coincides with the covariant \rep{}
$T$ defined in Subsection~$\ref{thep}$.
\end{teo}
\begin{beweis}
From Definition~\ref{YEmbed}, Lemma~\ref{Limit} and noting that
$J(g,Z)^{-1}=J(g^{-1},gZ)$ (use Eqs.~(\ref{J1}) and (\ref{J2}))
we have that
\[
 \lim\limits_{{\cal V}_+\ni gy\to 0}
\LE\; \int\limits_{\R^4}
  e^{-i\,\KIn{Tr$\left( P^\dagger (gZ)^* \right)$}} 
  J^{\KIn{$(0,\frac{n}{2})$}}(g^{-1},g(X+iY))^* f(x)\,\Wort{d}^4x \RE_+
 = \LE \left( \widehat{T_0(g)f}\right)(p)  \RE_+ .
\]
That $W_1\hrist\UPgr=V_1'$ follows by the same arguments 
as in the proof of Theorem~\ref{ExtW1}.
Note also that for suitable $f$ and $g$ as stated above the test function
$T_0(g)f$ is smooth and the support properties of $T_0$ follow immediately
from its definition. Finally, we have for elements
$g_0=\left(\kern-1.5mm \begin{array}{cc}
                      A \kern-2mm &  C (A^*)^{-1} \\
                      0 \kern-2mm &  (A^*)^{-1}
                 \end{array} \kern-1.5mm\right)\in \UPgr$
that $J^{\KIn{$(0,\frac{n}{2})$}}(g_0,Z)=D^{\KIn{$(0,\frac{n}{2})$}}
(A^{-1})^*$ and this implies $T_0\hrist\UPgr =T$
on the space of test functions, where the covariant
\rep{} $T$ is given in Eq.~(\ref{VfCov}).
\end{beweis}

\begin{rem}
Taking into account the comments in Remark~\ref{OtherSpaces}~(ii)
we can define similarly the \rep{s} $W_i$, 
$i=2,3,4$, and obtain the corresponding intertwining
relations with $T_0$.
\end{rem}

\begin{teo}\label{TConfCov}
The massless free nets given in Theorem~$\ref{Prop}$ are 
$\g{SU}{2,2}$ covariant.
\end{teo}
\begin{beweis}
Putting in the Bose case ($n$ even) $W_n:=W_1\oplus W_2$
and in the Fermi case ($n$ odd)
$W_n:=W_1\oplus W_2\oplus W_3\oplus W_4$, we get from
Theorem~\ref{DefT0} and the preceding remark 
that for $n$ even $W_n$ leaves the symplectic form $\sigma_n$ invariant 
resp.~for $n$ odd $W_n$ leaves the corresponding scalar product
on $h_n$ invariant (cf.~Definition~\ref{factornets}). Further,
for suitable $g\in\g{SU}{2,2}$ and 
$f\in\TestO{\R^{4},{\cal H}^{\KIn{$(0,\frac{n}{2})$}}}$
the respective eq.
\[
 W_n(g){\got I}_n f={\got I}_n(T_0(g)f)
\]
hold. Now the covariance follows by similar arguments as in the proof of
Theorem~\ref{Prop}.
\end{beweis}

We will now make use of the $\g{SU}{2,2}$ covariance proved in the
preceding theorem and which is typical of massless free nets.
We will show that the models studied in this paper
are examples of the conformally covariant nets studied in
\cite{Brunetti93}. Thus at the level of the von Neumann algebras 
we will be able to apply the general results of the mentioned  
reference. First we need to consider 
the natural Fock states on $\ccr{h_n,\sigma_n}$ 
resp.~$\car{h_n,\Gamma_n}$. For $n$ even the Fock state is
specified by the natural internal complexification 
of $h_n$, $j(\varphi_+\oplus
\varphi_-):=i\varphi_+\oplus i\varphi_-$, while for $n$ odd the Fock
state characterized by the basis projection 
$P:=  \left(\kern-1.5mm \begin{array}{cccc}
 \EINS \kern-1mm & 0 \kern-1mm & 0 \kern-1mm & 0            \\
   0 \kern-1mm & \EINS \kern-1mm & 0 \kern-1mm & 0           \\ 
   0 \kern-1mm & 0 \kern-1mm & 0 \kern-1mm & 0             \\
   0 \kern-1mm & 0 \kern-1mm & 0 \kern-1mm & 0                        
      \end{array} \kern-1.5mm\right)$ on $h_n$ 
(cf.~\cite[Chapter~8]{bBaumgaertel92},\cite{Lledo95}). Note that in 
both cases the one particle Hilbert space is given by
$h_n:={\got H}_{n,+}'\oplus {\got H}_{n,-}'$ and the unitary reducible
representation $W_1'\oplus W_2'$ satisfy the spectrality condition on
it (recall also Remark~\ref{OneP}). 
Let $\pi_0$ be the Fock \rep{} on the corresponding symmetric
resp.~antisymmetric Fock space ${\got H}_0$ with Fock vacuum
$\Omega$ and denote by a prime the commutant 
in $\al B.({\got H}_0)$. Then we may consider the 
following net of von Neumann algebras indexed by double cones:
\[
 \al K.\ni\al O.\mapsto\al M._n(\al O.)
   :=\LR \pi_0\LR\al A._n(\al O.)\RR \RR''\subset\al B.({\got H}_0)  .
\]

We will show next that the preceding net
$\al O.\mapsto\al M._n(\al O.)$ satisfies the axioms of a vacuum \rep{}
(cf.~\cite[Chapter~1]{bBaumgaertel95}) with the stronger covariance 
w.r.t.~the conformal group.

\begin{pro}
The nets of von Neumann algebras $\al O.\mapsto\al M._n(\al O.)$,
$n\in\N$, defined before satisfy the properties of 
\begin{itemize}
\item[{\rm (i)}] $($Isotony$)$
 If $\B_1\subseteq\B_2$, then
 $\al M._n(\B_1)\subseteq\al M._n(\B_2)$, $\al O._1,\al O._2\in\al K.$.
\item[{\rm (ii)}] $($Causality$)$
 If $\B_1\perp\B_2$, then $\al M._n(\B_1)\subseteq\al M._n(\B_2)'$.
\item[{\rm (iii)}] $($Additivity$)$
 For any $\{\al O._\lambda\}_{\lambda\in\Lambda}\subset\al K.$ with
 $\cup_\lambda\al O._\lambda\in\al K.$. Then 
\[
 \al M._n(\cup_\lambda\al O._\lambda)=\bigvee_\lambda 
   \al M._n(\al O._\lambda)
  :=\LR \cup_\lambda \al M._n(\al O._\lambda)\RR''.
\] 
\item[{\rm (iv)}] $($Covariance and spectrality condition$)$
There exists a unitary \rep{} $Q$ of $\g{SU}{2,2}$ on 
$\al B.({\got H}_0)$ and a $Q$--invariant vector $\Omega\in
{\got H}_0$ such that $\al M._n(g\B)=Q(g)\,\al M._n(\B)\,Q(g)^{-1}$,
$g\in\g{SU}{2,2}$. Further $Q\hrist\UPgr$ is strongly continuous
and the generators of the space time translations satisfy the
spectrality condition.
\end{itemize}
\end{pro}
\begin{beweis}
The properties of isotony, causality and additivity follow directly 
from the corresponding properties of the net of abstract
C*--algebras $\al O.\mapsto\al A._n(\al O.)$ in Theorem~\ref{Prop}.
Further recall that for the Bose resp.~the Fermi case the 
one--particle Hilbert space
associated to the natural Fock representations is
${\got H}_{n,+}'\oplus {\got H}_{n,-}'$ 
and since $W_1\oplus W_2$ given above is unitary on it we have from
the invariance of the Fock state and Theorem~\ref{TConfCov}
that for suitable $g\in\g{SU}{2,2}$ and $\al O.\in\al K.$
\[
 \al M._n(g\al O.)=\LR\pi_0\circ\alpha_g\left(\al M._n(\al O.)\right) \RR ''
    =\LR Q(g)\,\pi_0\left(\al M._n(\al O.)\right)\,Q(g)^{-1} \RR ''    
    =Q(g)\,\al M._n(\al O.)\,Q(g)^{-1}\,.
\]
Here $Q(g)$ is the second quantization of $W_1\oplus W_2$
on the symmetric resp.~antisymmetric Fock space over 
${\got H}_{n,+}'\oplus {\got H}_{n,-}'$. Further,
$Q(g)\,\Omega=\Omega$, where $\Omega$ is the Fock vacuum
and $Q\hrist\R^4$ satisfies the spectrum condition
because by Theorem~\ref{DefT0}
$(W_1\oplus W_2)\hrist\R^4 = 
(V_1'\oplus V_2')\hrist\R^4$ does.
\end{beweis}

For unbounded regions one defines the corresponding 
localized von Neumann algebras by additivity.

We will conclude this section mentioning some standard algebraic
results for these models that are consequence of the conformal 
covariance showed above. We will freely use in the following
definitions and results from \cite[Section~2]{Brunetti93},
\cite[Section~4]{Hislop88} and \cite{Hislop82} (cf.~also
with references cited therein). Denote by $K_1$ the
double cone of radius 1 and centered at the origin, by
$\al W._r:=\{x\in\R^4\mid |x_0|<x_3\}$ the right wedge and
by $\al V._+$ the forward light cone. Recall that there are
elements of the conformal group that map these regions in
each other. Then we have:
\begin{itemize}
\item[(i)] The von Neumann algebras $\al M.(K_1)$,
 $\al M.(\al W._r)$ and $\al M.(\al V._+)$ are spacially
 isomorphic and in particular Type~III$_1$--factors.
\item[(ii)] The modular groups of the von Neumann algebras 
 $\al M.(K_1)$, $\al M.(\al W._r)$ and $\al M.(\al V._+)$ act 
 geometrically.
\item[(iii)] Implementation of the PCT transformation using the
 modular conjugation associated to the von Neumann algebra 
 $\al M.(\al W.)$ for a wedge region $\al W.$.
\item[(iv)] Essential duality and timelike duality for 
 the forward/backward cones hold.
\end{itemize}

\section{Conclusions}
We have seen in this paper that the notion of free net (which avoids the 
explicit use of quantum fields) is particularly well adapted in the massless
case for proving standard properties expected for these models,
in particular for showing covariance under the conformal group.
Further, free nets are completely characterized by the embeddings
$\ot I._n$ (cf.~Section~\ref{mfn}) which have a purely group theoretical
interpretation. 

One possible extension of the massless free net construction is to
consider higher dimensional (flat) Minkowski space (although it is not
clear that this would be physically meaningful). In any case some of 
the important features of the construction presented here still appear in 
higher dimensions. Indeed, in a recent paper by Angelopoulos and Laoues 
 \cite{Angelopoulos98} with the suggestive title ``{\em Masslessness in 
 $n$--dimensions}'' it is shown that some of the characteristic group 
 theoretical aspects of the 4--dimensional theory are still valid for
 $n\geq 5$. In particular, the notion of massless representations (which
 are again induced representations) can be naturally stated in this context
 and it is still true that they extend to unitary representations of the 
 corresponding conformal group. A new aspect of higher dimensions though
 is the fact that the degeneracy of the inducing representations of the 
 associated little groups $\al E.(n-2)$ affects also its `rotational' part.
 Thus generalizing the notion of covariant representation to this situation
 we conclude that the reduction of the degrees of freedom mentioned in
 Sections~\ref{Intro} and \ref{mfn} will be even more present in
 higher dimensions.

\paragraph{Acknowledgements}
It is a pleasure to thank Sergio Doplicher and Roberto Longo
for their hospitality at the Mathematics Departments of the
universities of Rome `La Sapienza' and `Tor Vergata', respectively.
The visit was supported by a EU TMR network 
``Implementation of concept and methods from
Non--Commutative Geometry to Operator Algebras and its applications'', 
contract no.~ERB FMRX-CT 96-0073. I would also like to acknowledge
useful conversations with Roberto Longo and Wolfgang Junker.


\providecommand{\bysame}{\leavevmode\hbox to3em{\hrulefill}\thinspace}

\end{document}